\documentclass[]{spie}
\pdfoutput=1
\usepackage[dvips]{graphicx}    % required for `\includegraphics'(yatex added)
\usepackage{amsmath}    % required for `\cases' (yatex added)
\usepackage{amssymb}
\usepackage{color}
\usepackage{aas_macros}
\usepackage{subfig}
\title{CLASS: The Cosmology Large Angular Scale Surveyor}

\author{Thomas Essinger-Hileman,\supit{a}  Aamir Ali,\supit{a} Mandana Amiri,\supit{b} John W. Appel,\supit{a} Derek Araujo,\supit{c} Charles L. Bennett,\supit{a} Fletcher Boone,\supit{a} Manwei Chan,\supit{a} Hsiao-Mei Cho,\supit{d} David T. Chuss,\supit{e} Felipe Colazo,\supit{e} Erik Crowe,\supit{e} Kevin Denis,\supit{e} Rolando D\"unner,\supit{f} Joseph Eimer,\supit{a} Dominik Gothe,\supit{a} Mark Halpern,\supit{b} Kathleen Harrington,\supit{a} Gene Hilton,\supit{d} Gary F. Hinshaw,\supit{b} Caroline Huang,\supit{a} Kent Irwin,\supit{g} Glenn Jones,\supit{c} John Karakla,\supit{a} Alan J. Kogut,\supit{e} David Larson,\supit{a} Michele Limon,\supit{c} Lindsay Lowry,\supit{a} Tobias Marriage,\supit{a} Nicholas Mehrle,\supit{a} Amber D. Miller,\supit{c} Nathan Miller,\supit{e} Samuel H. Moseley,\supit{e} Giles Novak,\supit{h} Carl Reintsema,\supit{d} Karwan Rostem,\supit{a,e} Thomas Stevenson,\supit{e} Deborah Towner,\supit{e} Kongpop U-Yen,\supit{e} Emily Wagner,\supit{a} Duncan Watts,\supit{a} Edward Wollack,\supit{e} Zhilei Xu,\supit{a} and Lingzhen Zeng\supit{i}                                     
\skiplinehalf                                                                           
\supit{a}Dept. of Physics and Astronomy, Johns Hopkins University, Baltimore, MD 21218, USA; \\
\supit{b} Dept. of Physics and Astronomy, University of British Columbia, Vancouver, BC V6T 1Z4, Canada;\\
\supit{c}Dept. of Physics, Columbia University, New York, NY, 10027 USA;\\
\supit{d}National Institute of Standards and Technology, 325 Broadway, Boulder, CO 80305, USA;\\
\supit{e}Code 660, NASA Goddard Space Flight Center, Greenbelt, MD 20771, USA;\\
\supit{f}Instituto de Astrofisica, Pontificia Universidad Católica de Chile, Santiago, Chile;\\
\supit{g}Dept. of Physics, Stanford University, Stanford, CA 94305, USA;\\
\supit{h}Dept. of Physics and Astronomy, Northwestern University, 2145 Sheridan Road, Evanston, IL 60208-3112, USA;\\   
\supit{i}Harvard-Smithsonian Center for Astrophysics, Cambridge, MA 02138, USA\\
}

\authorinfo{Further author information: (Send correspondence to T. Essinger-Hileman)\\ E-mail: essinger@pha.jhu.edu}

\begin{document}

\maketitle

\begin{abstract}
The Cosmology Large Angular Scale Surveyor (CLASS) is an experiment to measure the signature of a gravita-tional-wave background from inflation in the polarization of the cosmic microwave background (CMB). CLASS is a multi-frequency array of four telescopes operating from a high-altitude site in the Atacama Desert in Chile. CLASS will survey 70\% of the sky in four frequency bands centered at 38, 93, 148, and 217 GHz, which are chosen to straddle the Galactic-foreground minimum while avoiding strong atmospheric emission lines. This broad frequency coverage ensures that CLASS can distinguish Galactic emission from the CMB. The sky fraction of the CLASS survey will allow the full shape of the primordial B-mode power spectrum to be characterized, including the signal from reionization at low $\ell$. Its unique combination of large sky coverage, control of systematic errors, and high sensitivity will allow CLASS to measure or place upper limits on the tensor-to-scalar ratio at a level of $r=0.01$ and make a cosmic-variance-limited measurement of the optical depth to the surface of last scattering, $\tau$.
\end{abstract}

\section{Introduction}
An inflationary phase in the first moments after the Big Bang has been used to explain the remarkable homogeneity, isotropy, and flatness of our universe, as well as the origin of the density perturbations that seeded large-scale structure. Inflationary theories naturally produce a spectrum of nearly scale-invariant, gaussian perturbations that have both a scalar component, which gives rise to perturbations in particle and radiation densities, and a tensor component, producing a gravitational-wave background. Inflation correctly predicted the existence of a slightly red initial spectrum of density perturbations.\cite{2013ApJS..208...20B, 2013arXiv1303.5076P, 2014JCAP...04..014D, 2012ApJ...755...70R} 

Recently, the BICEP2 experiment announced the detection of $B$-mode polarization at $\ell$ of 40-200,~\cite{2014PRL1403.3985B} but it is unclear whether this signal is cosmological or Galactic in nature. These results have generated strong interest in complementary experiments and have highlighted the importance of multi-frequency observations for foreground subtraction. A measurement of $B$-modes in the CMB would constitute important evidence for inflation and a measurement of the energy scale at which inflation occured. The tensor-to-scalar ratios, $r \leq 0.1$, being probed correspond to $E \sim 10^{16}$ GeV, near grand-unified-theory (GUT) energy scales. The gravitational waves from inflation are our only probe of the physics at such enormous energies and at such early times, just $10^{-35}$ seconds after the Big Bang. They would also provide the first firm evidence for the existence of quantum-gravitational effects.\cite{2014PhRvD..89d7501K} Detecting primordial gravitational waves requires greater frequency coverage to definitively rule out Galactic foreground contamination, as well as a measurement of the $B$-mode signal over a wider range of angular scales to verify the full shape of the $B$-mode power spectrum.

A number of experiments are searching for $B$-mode polarization. Notably, the Planck satellite has mapped the entire sky in nine frequency bands from 30 to 857 GHz, allowing measurement of CMB polarization over a broad range of angular scales with the ability to remove Galactic foreground contamination; however, it is yet to be seen whether Planck will have the ability to constrain this signal. In this paper we present the Cosmology Large Angular Scale Surveyor (CLASS), which is leading the effort to map the CMB polarization at large angular scales from the ground. CLASS will observe in four frequency bands centered on 38, 93, 148, and 217 GHz. CLASS is uniquely poised to measure inflationary gravitational waves through its ability to measure CMB polarization at the largest angular scales, and has the power to discriminate polarized Galactic emission from CMB polarization due to its broad frequency coverage straddling the Galactic foreground minimum. CLASS will also make cosmic-variance-limited measurements of $E$-mode polarization for angular scales below $\ell \sim 100$, allowing a measurement of the optical depth to reionization with 4.5\% error, an improvement of 35\% over Planck. CLASS will operate from a high-altitude site in the Atacama Desert of Chile, with the Q-band, 38 GHz, telescope deploying first in early 2015.

This paper is organized as follows: Section~\ref{sec:measurement} provides an overview of the status of CMB polarization measurements and the expected $B$-mode signal. Section~\ref{sec:strategy} introduces the CLASS experiment, its observational strategy, and its expected sensitivity. Sections~\ref{sec:detectors} and~\ref{sec:receivers} describe the detectors and dilution-refrigerator receivers, respectively, employed by CLASS. Section~\ref{sec:optics} outlines the optical elements of CLASS, while Section~\ref{sec:vpm} provides details of the polarization modulators used by CLASS. The telescope mounts, Atacama site, and deployment plan are laid out in Section~\ref{sec:mount_and_site}. Finally, Sections~\ref{sec:qband_receiver} and~\ref{sec:wband_receiver} describe the unique aspects of the Q- and W-band instruments, primarily focusing on the focal-plane design.

\section{The Search for Primordial $B$-Modes}
\label{sec:measurement}
The CMB is a particularly clean probe of the state of the early universe. Approximately 90\% of CMB photons were last scattered during a narrow window in time, termed recombination, around 400,000 years after the Big Bang and have free-streamed for the intervening 13.8 billion years. The remaining 10\% of photons scattered at late times during reionization, providing additional constraints on the epoch of reionization. Temperature anisotropy is observed in the 2.725 K\cite{1999ApJ...512..511M} CMB at an RMS level of 1 part in 100,000, or approximately 50 $\mu$K. Observations of the temperature anisotropy are a probe of the density perturbations at decoupling and have been instrumental in placing constraints on parameters of $\Lambda$CDM cosmology at the 1\% level.\cite{2013ApJS..208...20B, 2013arXiv1303.5076P, 2014JCAP...04..014D, 2012ApJ...755...70R}

In addition to temperature anisotropy, the CMB is also faintly polarized at less than 1 part-per-million with roughly 1 $\mu$K RMS amplitude. The polarization in the CMB is produced by inverse Thomson scattering, and has been observed by a number of experiments to date.\cite{2010ApJ...711.1123C, 2006ApJ...647..813M, 2008ApJ...684..771B, 2007ApJ...660..976S, 2005ApJ...624...10L, 2009ApJ...705..978B, 2007ApJ...665...55W, 2012arXiv1207.5034Q, 2013ApJS..208...20B, 2014PRL1403.3985B, 2013arXiv1312.6646P, 2013PhRvL.111n1301H} Both scalar and tensor perturbations produce polarization, but these two sources of CMB polarization can be separated by performing a decomposition into non-local divergence and curl terms, referred to as $E$-modes and $B$-modes, respectively.

\begin{figure}[tbp]
\centering
  \includegraphics[width=0.7\textwidth]{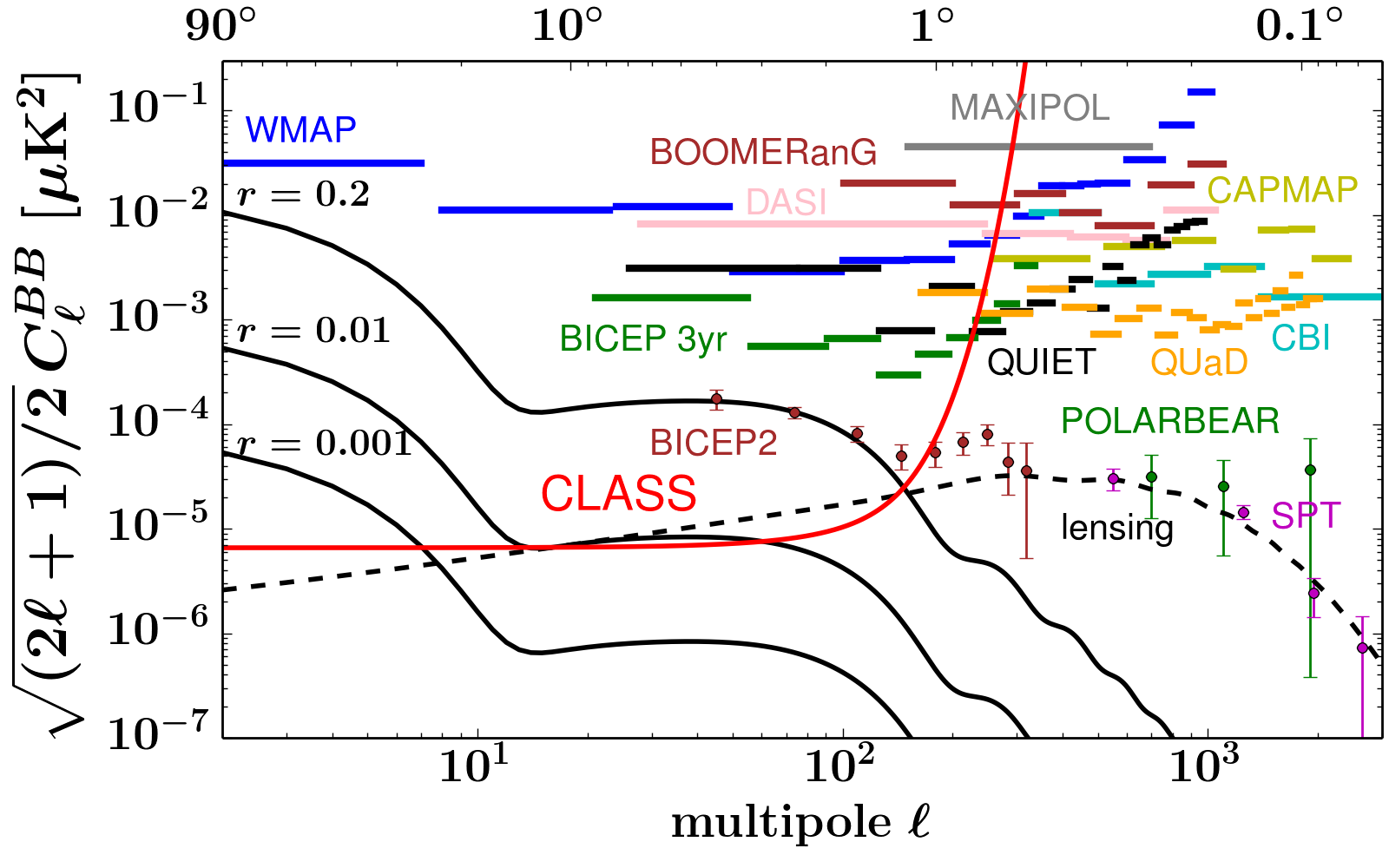}
  \caption{Primordial $B$-mode angular power spectra for $r$ values of 0.001, 0.01, and 0.2 are shown, along with the lensed $B$-mode power spectrum and quoted upper limits from the BICEP\cite{2010ApJ...711.1123C}, BOOMERanG\cite{2006ApJ...647..813M}, CAPMAP\cite{2008ApJ...684..771B}, CBI\cite{2007ApJ...660..976S}, DASI\cite{2005ApJ...624...10L}, QUaD\cite{2009ApJ...705..978B}, MAXIPOL\cite{2007ApJ...665...55W}, QUIET\cite{2012arXiv1207.5034Q}, and WMAP\cite{2013ApJS..208...20B} experiments. The measured $B$-mode spectra from the BICEP2\cite{2014PRL1403.3985B}, POLARBEAR\cite{2014arXiv1403.2369T}, and SPT\cite{2013PhRvL.111n1301H} experiments are also shown with 1-$\sigma$ error bars. The full-survey, four-frequency CLASS sensitivity is shown in red.}
\label{fig:class_sensitivity}
\end{figure}

Scalar perturbations only produce $E$-modes, while tensor modes produce both E- and $B$-modes. Measuring primordial $B$-mode polarization in the CMB thus constitutes a detection of inflationary gravitational waves. The expected angular power spectrum of the primordial $B$-modes is shown in Figure~\ref{fig:class_sensitivity} along with the current state of $B$-mode measurements. The $B$-mode signal from decoupling is greatest at the ``recombination peak'' around $\ell=80$, which is the targeted angular scale for most ground-based observatories. An even stronger $B$-mode signal is expected at $\ell < 15$ from photons scattering off free electrons during reionization. CLASS is designed to capture the full shape of the primordial $B$-mode power spectrum, including both this ``reionization bump'' and the recombination peak.

To detect inflationary gravitational waves, experiments must separate primordial $B$-mode polarization from two other sources of $B$-modes: gravitational lensing of $E$-modes into $B$-modes and polarized Galactic emission from dust and synchrotron sources. The expected lensed $B$-mode power spectrum is shown as the dashed line in Figure ~\ref{fig:class_sensitivity}. The lensing signal is expected to peak at smaller angular scales, $\ell \sim 700$. From the figure it is clear that even if $r=0.001$ the primordial $B$-modes are observable above the lensed $B$-modes at the largest angular scales probed by CLASS.

Polarized Galactic foregrounds are a significant contaminant for $B$-mode experiments and are to date poorly constrained at the levels needed for understanding their effects on the recovery of $r$. Synchrotron emission from relativistic electrons accelerated by Galactic magnetic fields is the dominant emission at low frequencies with a sharply-falling spectrum with average index around -3 in units of antenna temperature. Polarized thermal emission from dust grains aligned with magnetic fields is dominant for higher frequencies and has an average spectral index around 1.8. Both dust and synchrotron emission show some indication of having varying spectral index on the sky.~\cite{2013arXiv1312.1300P, 2014arXiv1404.5323F} These two foregrounds reach a minimum around 65 GHz,\cite{2007ApJ...665..355K} as indicated by WMAP.~\cite{2013ApJS..208...20B} Planck has recently released maps of polarized emission for a fraction of the sky, which indicates that dust emission can have polarization fractions up to 18\%, particularly in regions with low dust emission intensity.~\cite{2014arXiv1405.0871P} Upcoming data releases from the Planck satellite will provide tighter constraints on the polarization fraction of foregrounds.

\begin{figure}[tbp]
\centering
  \includegraphics[width=0.6\textwidth]{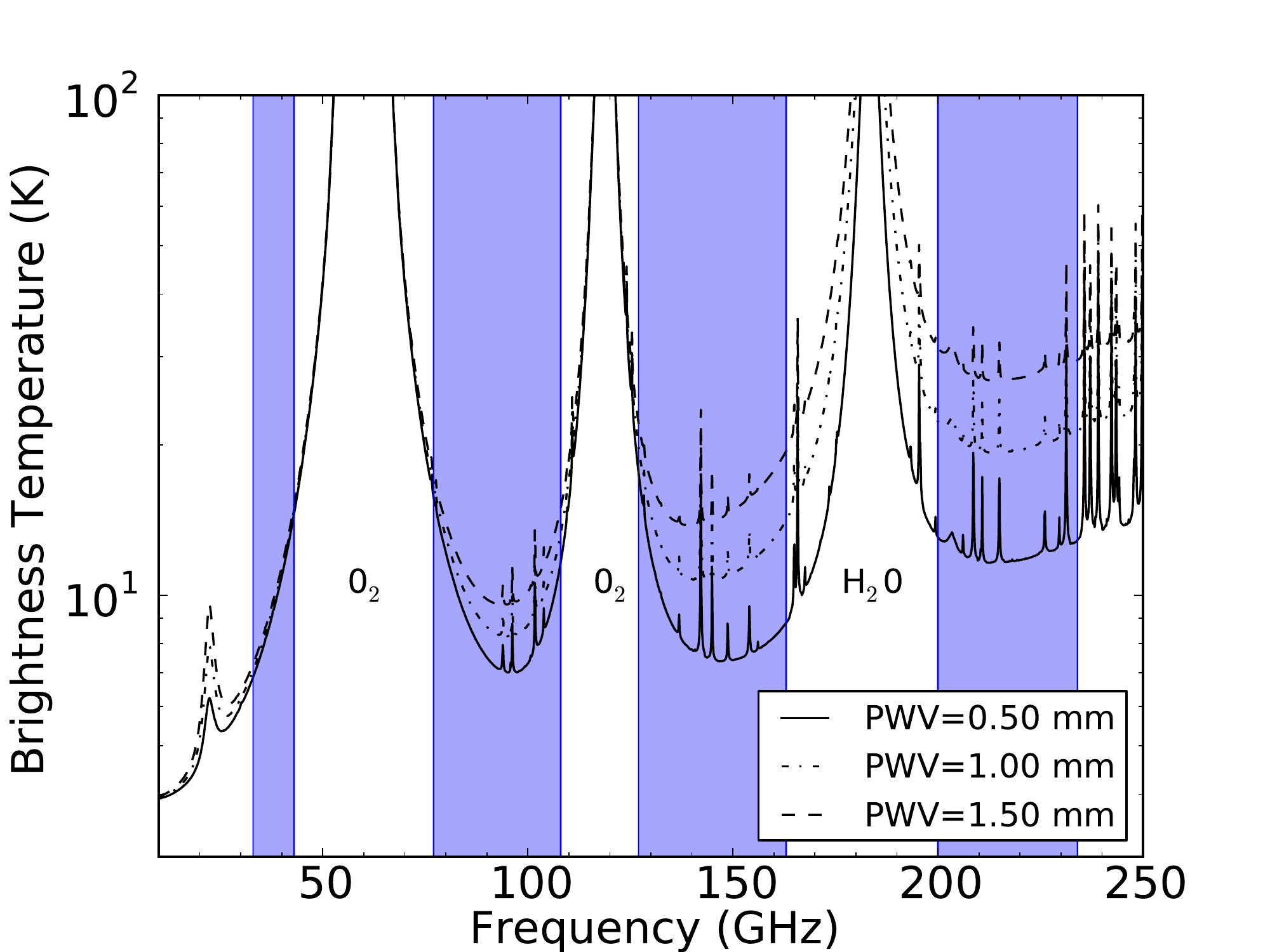}
  \caption{Estimated mean atmospheric brightness temperature (Rayleigh-Jeans) versus frequency for precipitable water vapor (PWV) levels of 0.5, 1.0, and 1.5 mm along with the CLASS bandpasses in blue, for an atmosphere at physical temperature of 250 K.~\cite{Pardo_atmosphere_atacama} The CLASS bands were chosen to avoid the prominent oxygen and water emission lines. As PWV increases, the broad water line at 183 GHz contaminates the 93, 148, and 217 GHz bands in particular. Median PWV in the Atacama is around 1.3 mm.\** }%\footnotemark}
\label{fig:class_bands}
\end{figure}

Though the Atacama Desert is one of the premier sites in the world for millimeter-wave astronomy, the dominant optical loading and noise source for CLASS is still the atmosphere. A frequency spectrum of the atmosphere is shown in Figure~\ref{fig:class_bands}. The strong oxygen and water lines seen there preclude observations except in discrete bands between them. The CLASS experiment will observe in four such bands centered at 38, 93, 148, and 217 GHz. The band edges are chosen to maximize sensitivity to the CMB versus noise from atmospheric loading. The atmospheric loading for a dry atmosphere and a loading per mm precipitable water vapor (PWV) for the CLASS bands are summarized in Table~\ref{tbl:class_instrument}. The loading for the yearly median PWV around 1.3 is estimated to be 0.9, 2.6, 3.8, and 6.6 pW for the 38, 93, 148, and 217 GHz CLASS bands, respectively. These numbers include estimated optical efficiencies for each band.%\footnote{These numbers include estimated optical efficiencies for the observing bands.}

Additionally, the atmosphere has a complicated and time-dependent structure, consistent with a turbulent layer with inhomogeneously-distributed water vapor with a physical distribution that is roughly consistent with a Kolmogorov model.~\cite{2000ApJ...543..787L, 2010ApJ...708.1674S, 2005ApJ...622.1343B} The atmosphere thus injects low-frequency (``$1/f$'') correlated noise into detector timestreams. The RMS level of atmospheric fluctuations from this turbulent layer has been estimated to have a mean value of 34 mK at 40 GHz.\cite{2000ApJ...543..787L} We can scale this to the other frequency bands using the loading versus PWV estimates in Table~\ref{tbl:class_instrument}, which gives 0.54, 1.73, and 3.73 K RMS fluctuations at 93, 148, and 217 GHz, respectively. This assumes that the same turbulent layer of water vapor is responsible for atmospheric $1/f$ noise in each band, and that other sources of loading such as oxygen are uniformly distributed.

\begin{figure}[t]
\centering
\subfloat[]{
		\includegraphics[height=0.45\textwidth, clip=true, trim=0in 0in 0in 0in]{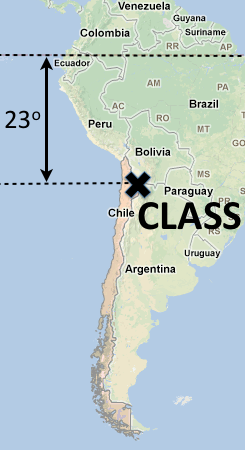}
}
\subfloat[]{
		\includegraphics[clip=true, trim=1in 0 1in 0, angle=0, height=0.45\textwidth]{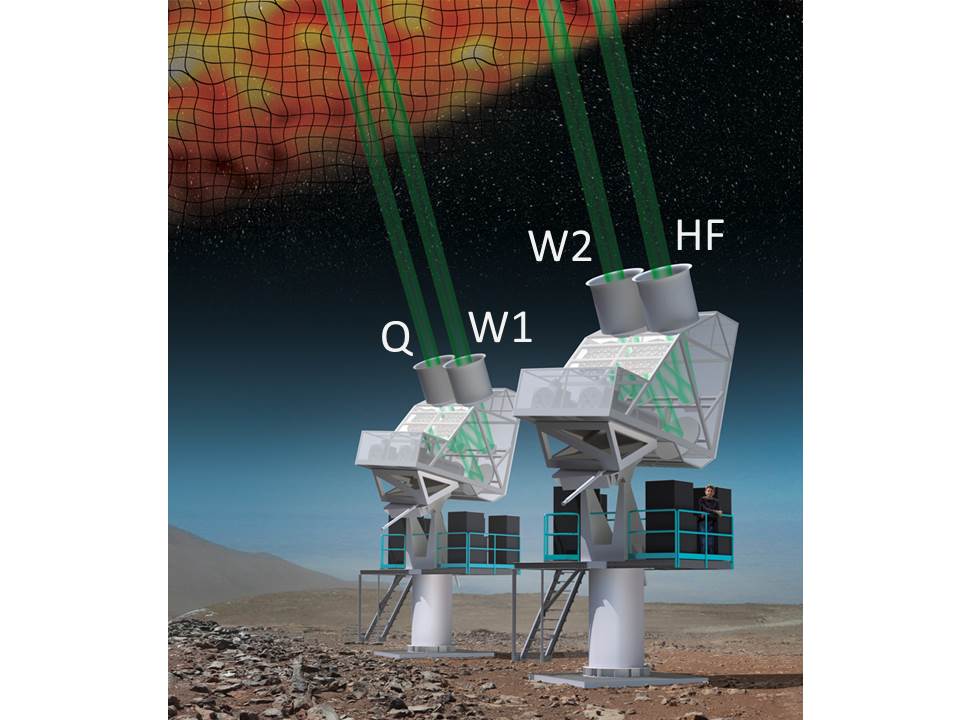}
\label{fig:class_mockup}}
\caption{(a) CLASS is located in the Atacama Desert of Chile near the equator at a latitude of approximately -23$^{\circ}$, allowing 70\% of the sky to be surveyed at 45$^{\circ}$ elevation. (b) An artist's conception of the CLASS array, with four telescopes on two three-axis mounts. Green lines show the optical path. The arrangement of the four frequency receivers is also shown.}
\label{fig:class_site}
\end{figure}

\section{Experimental strategy}
\label{sec:strategy}
The CLASS experiment is optimized to measure CMB polarization over the angular range where the primordial $B$-mode signal could dominate over the lensed signal, $\ell \sim 5$--$100$. This makes it capable of verifying the full shape of the $B$-mode power spectrum, and in particular of capturing the reionization bump at low $\ell$, which is required for full confirmation of a discovery of inflationary gravitational waves. Additionally, CLASS will have the ability to reject foreground contamination with high fidelity due to its broad frequency coverage that emphasizes measurements surrounding the foreground minimum. CLASS will deploy one Q-band (38 GHz) receiver; two W-band (93 GHz) receivers; and a high-frequency dichroic receiver with two bands centered at 148 and 217 GHz. Hereafter, these instruments will be referred to as the Q, W1, W2, and HF instruments, with HF150 and HF220 referring to the individual bands of the high-frequency instrument when necessary.

To achieve its goals, CLASS must have high sensitivity, control over sources of systematic errors, and stability in its measurement of polarization over long time scales. These requirements drive the design of the CLASS instruments, including their choice of detector technology, low detector base temperature provided by dilution refrigerators, optical layout, fast polarization modulation using a variable-delay polarization modulator (VPM), co-moving ground shield, three-axis (azimuth, elevation, and boresight) mounts, and observing site in the Atacama Desert of Chile.

The four telescopes of CLASS share much of their design in common, allowing early validation of many elements on the Q telescope with the possibility of iteration for later telescopes. This facilitates rapid deployment of multiple telescopes with a minimum of additional work, as well as mitigating the risk associated with a given design.

\footnotetext{\** APEX weather monitor, www.apex-telescope.org/weather/ }

\subsection{Sensitivity}
The sensitivity of the CLASS telescopes, summarized in Table \ref{tbl:class_instrument}, is achieved through a combination of highly efficient and reproducible detectors, high-efficiency optics, choice of W-band-centered frequency coverage, and transition-edge-sensor (TES) bolometers cooled to 70 mK by a dilution refrigerator. The CLASS sensitivity is estimated from a combination of laboratory measurements of detector parameters and physical optics simulations. Three primary sources of noise are considered in estimating the CLASS sensitivity: detector noise, SQUID amplifier noise in the multiplexing readout, and photon noise intrinsic to the measured optical power detected. %Details of the sensitivity calculation are given in Appendix \ref{app:sensitivity}. 

The CLASS detectors, described in Section \ref{sec:detectors}, achieve optical efficiency between the OMT probes and the transition-edge-sensor (TES) bolometer of 90\% at Q band,\cite{Rostem2012SPIE} with similar efficiency expected at the other frequency channels. This high detector efficiency, when combined with high-efficiency optical elements and low cold-stop edge taper, allows the CLASS system to have an estimated 68\% system optical efficiency in the Q band and 56\% optical efficiency at W band. Table~\ref{tbl:qband_loading} shows the details of the optical efficiency calculation for the Q telescope. An experiment with 70\% detector efficiency (40\% total efficiency) but otherwise identical to CLASS would be approximately 1/3 less sensitive.

The 70 mK base temperature of the CLASS dilution refrigerator, compared to the 300 mK base temperatures achievable with helium adsorption refrigerators, also reduces detector phonon noise. The bolometer phonon noise scales as $n_{d} \propto T_{c}$ for a TES superconducting transition temperature, $T_{c}$. CLASS detectors with $T_{c} \sim 150$ mK, compared to 500 mK transition temperatures for adsorption-fridge systems, have approximately 50\% lower detector noise. Especially at Q band where total optical loading and noise are low, this decrease in detector noise has a significant impact on total system noise.

Finally, the choice to have two W-band telescopes for CLASS maximizes the CMB-to-foreground ratio at large angular scales. The atmospheric loading at 90 GHz is lower than at 150 GHz, making the per-detector W-band sensitivity 50\% better than at 150 GHz. The combination of this with high optical efficiency and low detector temperature means that a single CLASS W detector can be twice as sensitive to the CMB as a 150 GHz detector with average optical efficiency operating at 300 mK base temperature. Because array sensitivity goes as the detector sensitivity times the square root of the number of detectors, the CLASS W-band telescopes with 259 detector pairs each are equivalent in sensitivity to kilopixel arrays of detector pairs.

Given the estimated detector sensitivities, $\sigma_{d}$, in Table \ref{tbl:class_instrument}, one can calculate projected array sensitivities as $\sigma_{a} = \sigma_{d} (\alpha N_{det})^{-1/2}$, where $\alpha$ is the average instantaneous fraction of working detectors. Many factors reduce $\alpha$, including fabrication errors that permanently disable a detector, weather or fabrication variation that reduce the number of detectors that are biased on transition at any one time, and a variety of post-processing detector cuts. We take $\alpha$ to be 90\% for the Q array, in which individual detector chips can be screened before assembly in the focal plane, and 80\% for the monolithic detector wafers of the W and HF instruments. 

A map sensitivity can be calculated from the instantaneous array sensitivity, given an observed area, and total observing time $\beta t$, where the observing efficiency $\beta$ is the fraction of time where the telescope was able to observe. Total observing time is reduced by periods of bad weather and equipment maintenance or repair, as well as cryogenic recycling for many experiments. This last inefficiency is eliminated for CLASS by using continuous dilution refrigerators. We take $\beta$ to be 50\% for all sensitivity calculations.

We assume that the Q and W1 receivers observe for 5 years, the W2 receiver observes for 3.5 years, and the HF receiver observes for 3 years. If we assume that the W1, W2, and HF150 receivers are used for CMB analysis, while the Q and HF220 channels are used for foreground cleaning, the final Q and U map sensitivity, combining all observations, would be 7 $\mu$K-arcmin over 70\% of the sky. This would be the deepest map of the CMB polarization at large angular scales to date, at least five times deeper than the Planck 143 GHz total survey sensitivity.

\begin{figure}[tbp]
\centering
  \includegraphics[width=0.7\textwidth]{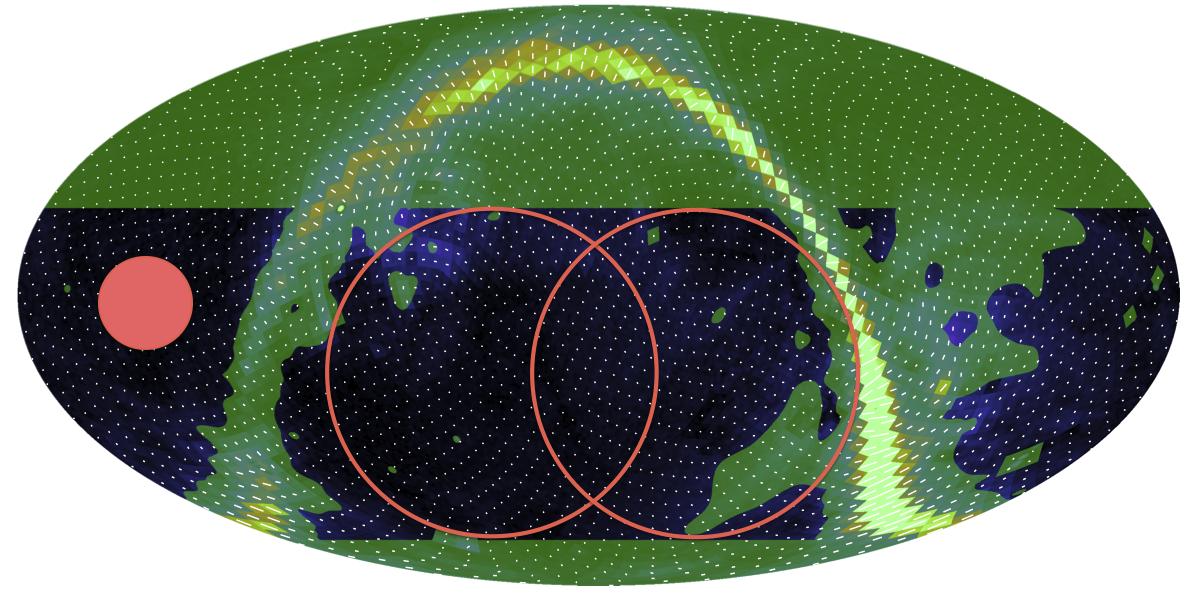}
  \caption{The CLASS survey covers 70\% of the sky. In this map in equatorial coordinates, the CLASS sky coverage is overlaid on a model of the polarized sky at 90 GHz. The green region will be masked for the CMB analysis and includes regions that CLASS will not observe as well as regions of strong Galactic emission. The planned 360$^{\circ}$ constant-elevation scans trace out the large red circles on the map. The filled red circle gives the instantaneous field of view of a CLASS telescope.}
\label{fig:class_survey}
\end{figure}

\subsection{Control of Systematic Errors}
In measuring small signals in the presence of much larger backgrounds, CMB polarization experiments need to be optimized to allow careful control and estimation of systematic errors. Contaminants to the polarization measurement include instrumental polarization, polarized ground pickup, and far-sidelobe response, as well as intensity-to-polarization leakage from beam asymmetries or imperfections in the VPM. Given that the intensity signal from the atmosphere and the CMB are both orders of magnitude above the polarized signal of interest, these effects need to be controlled carefully.

In addition to the many jack-knife tests that can be performed by splitting the data into segments and comparing maps, the CLASS instrument is designed to mitigate systematic uncertainties particularly through its use of the VPM and boresight rotation axis. The VPM modulates celestial polarization but not instrumental polarization, giving the CLASS instrument immunity from contamination from birefringence of dielectrics or polarization induced by reflections. The VPM also provides a means of separating unpolarized atmospheric emission from CMB polarization, reducing the most significant source of $1/f$ noise in the detector timestreams. This is critical in recovering large-scale CMB polarization from the ground. Additionally, the VPM provides a measurement of circular polarization. Because the CMB is not expected to be circularly polarized, this allows the V maps from CLASS, which should be null maps, to be powerful constraints on systematic uncertainties. Finally, the boresight-rotation axis allows the polarization angles of the CLASS instrument to be changed relative to the horizon, and will allow discrimination of polarization from ground pickup versus celestial polarization.

\subsubsection{Simulations}
To probe the ability of CLASS to recover $r$ in the presence of systematic errors, we have performed end-to-end simulations of CLASS observations incorporating various systematic effects that may contaminate the recovered CMB polarization angular power spectra. Simulated detector timestreams are generated for CLASS scanning a simulated CMB sky, with noise and systematic errors injected into the timestreams. The recovered maps can then be compared to the input maps to estimate the level of systematic errors for CLASS. Effects included in the modeling are errors in the assumed VPM transfer function from misalignment of the VPM and detector axes, temperature variation of the VPM, and detector gain variation. In the simulations, CLASS was made to scan the sky at constant elevation in large circles as in Figure \ref{fig:class_survey} with boresight rotation executed daily by 15$^{\circ}$. We found that by properly modeling each of these effects, one can recover high-fidelity maps of the Stokes Q and U parameters across 70\% of the sky. The resulting power spectra are free of systematics at a level more than two orders of magnitude below a $B$-mode signal for $r=0.01$. This will be treated in detail in an upcoming publication. (N. Miller, et al., 2014, in preparation) We continue to investigate other possible sources of systematic error, including a modulated VPM emission waveform from the maps. For more details of the VPM transfer functions, see Section~\ref{sec:vpm}.

\begin{table}[t]
\begin{minipage}[t]{\textwidth}
\centering
\begin{tabular}{|c|c|c|c|c|c|c|c|c|c|}
\hline
    Telescope & Band & Beam & Dry Atm. & H$_{2}$O Loading & $N_{TES}$ & TES & Array& Obs. & Survey \\
              &   & FWHM & Loading   &   &  & NEQ\footnote{Incorporates all efficiency factors in the system, including in Stokes Q measurements due to VPM modulation.} & NEQ\footnote{Array efficiency is assumed to be 90\% for Q telescope and 80\% for all others.}    & Time\footnote{Operation is 24 hr/day, with 50\% observing efficiency assumed.} & NEQ/U  \\
    &(GHz)& (arcmin) & (pW) & (pW$ / $mm PWV) & & ($\mu$K-$\sqrt{s}$) & ($\mu$K-$\sqrt{s}$) & (years) & ($\mu$K-$^{\prime}$) \\
\hline \hline
    Q     & 33-43  & 90 & 1.32 & 0.068 & 72  & 207 & 25 & 5.0 & 39 \\
    \hline
    W1    & 77-108 & 40 & 3.34 & 1.07 & 518 & 171 & 8.5 & 5.0 & 13 \\
    \hline    
    W2    & 77-108 & 40 & 3.34 & 1.07 & 518 & 171 & 8.5 & 3.5 & 16 \\
    \hline
    HF150\footnote{Values for the CLASS HF channel are preliminary.} & 127-163 & 24 & 2.74 & 3.45 & 2000 & 278 & 7  & 3.0 & 15 \\
    HF220 & 200-234 & 18 &  2.11 & 7.45 & 2000 & 820 & 21 & 3.0 & 43 \\
   	\hline
   	Total & -- & -- & -- & -- & 5108 & --  & 5  & 19.5 & 10 \\
   	\hline
\end{tabular}
\caption{CLASS key parameters}
\label{tbl:class_instrument}
\end{minipage}
\end{table}

\subsection{Scan Strategy}
The CLASS nominal observing strategy is to perform constant-elevation scans over 360$^{\circ}$ in azimuth at a speed $\sim 1^{\circ}$ per second. This scan strategy covers roughly 70\% of the sky every day, with the exception of an avoidance region around the sun. Boresight rotation will be performed once per day at approximately 15$^{\circ}$ increments to change the polarization direction on the sky from day to day. This is important in separating out polarized ground pickup from celestial polarization. Figure~\ref{fig:class_survey} shows the resulting sky coverage achievable with such scans at a boresight elevation of 45$^{\circ}$, along with a likely Galactic mask for the CMB analysis. Observing from norther Chile at a latitude of $-23^{\circ}$ is important for enabling coverage of such a large fraction of the sky. Observations from the South Pole at 45$^{\circ}$ elevation can only cover approximately 25\% of the sky, and would not be capable of capturing the reionization bump at low $\ell$. 

\subsection{Foreground removal}
\label{sec:foregrounds}
Given an estimated final map noise, we can simulate the likelihood of detecting $B$-modes in the presence of synchrotron and dust foregrounds, as well as a Galactic mask. We use a pixel-based likelihood to simultaneously fit for foreground amplitudes and spectral indices, along with a foreground-cleaned map.~\cite{2011ApJ...737...78K, 2009MNRAS.397.1355E} The exact likelihood encodes all covariance between E- and $B$-modes (including effects of the survey boundary and Galactic mask), so misestimation of $B$-modes due to ``E-B mixing'' is not an issue. We have done a Markov Chain Monte Carlo analysis for realizations of the CMB cut by our Galactic mask for various input values of $r$ and map noise in the range 5-20 $\mu$K-arcmin expected for CLASS. This can be computed efficiently for low $\ell$; our analysis has thus far focused on $\ell<23$. 

We assume only CLASS data is used in the foreground removal, though external data sets could be incorporated if needed. The Q map is used to clean synchrotron and the HF220 map is used to clean dust. The W1, W2, and HF150 maps are the primary CMB maps. Template-based foreground removal adds noise to the cleaned map in proportion to the input map's noise referenced to the central frequency through the relevant spectral index ($\sim$ $-3$ for synchrotron and $\sim$ 1.8 for dust). The scaling from the Q and HF220 channels to the CMB channel is taken to be 0.07, and 0.10, respectively. Given the survey sensitivities expected from Table~\ref{tbl:class_instrument}, the final cleaned CMB maps are estimated to have noise on the order of 10 $\mu$K-arcmin.

These simulations, to be presented in more detail in an upcoming paper,~\cite{watts_class_foregrounds2014} show that CLASS can recover $r=0.01$ at 95\% confidence in the presence of foregrounds with the estimated final map noise of 10 $\mu$K-arcmin. The inclusion of varying spectral indices for dust and synchrotron will degrade this performance, while the inclusion of higher $\ell$ modes in the analysis will tighten constraints. An upcoming publication will detail the CLASS foreground removal simulations. (D. Watts, et al., 2014, in preparation)

\begin{table}[t]
\begin{minipage}[t]{\textwidth}
\centering
\begin{tabular}{|c|c|c|c|c|c|}
\hline
  Element  & Temp & Emis       & Spill\footnote{Beam spill number also incorporates reflection.} &  Cumul. & Loading\footnote{Power (single mode, single polarization) on a detector from an element, including all efficiency factors.} \\
           & (K)  &  &      &  Effic.\footnote{Total efficiency from the detector to the element.} & (pW)      \\
\hline \hline
Detector   & 0.1 & ---    & ---   & 0.90 & ---   \\
Lens 1     & 0.1 & 0.01   & 0.003 & 0.89 & 0.001 \\ 
Nylon      & 3   & 0.04   & 0.015 & 0.84 & 0.017 \\
Lens 2     & 4   & 0.015  & 0.023 & 0.80 & 0.014 \\
Cold Stop  & 4   & 0.00   & 0.056 & 0.76 & 0.020   \\
PTFE 1     & 10  & 0.008  & 0.015 & 0.74 & 0.013 \\
PTFE 2     & 70  & 0.008  & 0.0008& 0.73 & 0.070 \\
PTFE 3     & 150 & 0.008  & 0.0009& 0.72 & 0.130  \\
Window     & 290 & 0.002  & 0.0003& 0.72 & 0.108 \\
Secondary  & 290 & 0.0002 & 0.0018& 0.72 & 0.055 \\
Primary    & 290 & 0.0002 & 0.0053& 0.72 & 0.155  \\
VPM        & 290 & 0.0002 & 0.0013& 0.71 & 0.041 \\
Atmosphere & 250 & 0.032  & ---   & 0.71 & 0.920 \\
CMB        & 2.7 & 1.00   & ---   & 0.68 & 0.181  \\
\hline
Total      &     &        &       &      & 1.72   \\
\hline
\end{tabular}
\caption{Optical loading from elements in the Q band receiver.}
\label{tbl:qband_loading}
\end{minipage}
\end{table}

\section{Detectors}
\label{sec:detectors}
The detectors for all CLASS receivers are produced at the Goddard Space Flight Center (GSFC).~\cite{2009AIPC.1185..371D} They are feedhorn-coupled, transition-edge-sensor (TES) bolometers, which use a planar membrane ortho-mode transducer (OMT) at the base of the feedhorn waveguide to couple two orthogonal linear polarizations over microstrip transmission lines to separate TES bolometers. The low base temperature and high efficiency of the CLASS detectors provides high per-pixel sensitivity. The TES detectors, operating from a base temperature around 70 mK, provided by a dilution refrigerator (see Section~\ref{sec:receivers}), have superconducting transition temperatures around 150 mK. This low base temperature means that the CLASS detectors have extremely low noise, making them fully background-limited at all but the Q band, where optical loading from the atmosphere is particularly low.

\begin{table}[h!]
\centering
\begin{tabular}{|c|c|c|c|c|c|c|c|}
\hline
 Parameter & T$_{c}$ & T$_{b}$ & R$_{n}$ & R$_{sh}$ & G at T$_{c}$ & P$_{sat}$ & NEP$_{d}$\\
\hline
 Value     & 156 mK & 70 mK & 10 m$\Omega$ & 250 $\mu \Omega$ & 184 pW/K & 6.8 pW & 11 aW$\sqrt{s}$\\
\hline
\end{tabular}
\caption{Q-band detector parameters, measured in test setups at JHU and GSFC and described in greater detail in companion proceedings.~\cite{appel_spie_2014,rostem_spie_2014}}
\label{tbl:qband_detectors}
\end{table}

The design and performance of the detectors and their implementation in the Q-band focal plane are described in greater detail elsewhere in these proceedings.~\cite{appel_spie_2014,rostem_spie_2014} Key parameters for the Q-band detectors are summarized in Table~\ref{tbl:qband_detectors}. These detectors have been measured to have typical noise levels of 11 aW $\sqrt{s}$. The bolometers are voltage biased by a $\sim 250$ $\mu \Omega$ shunt resistor and noise-bandwidth limited by a Nyquist inductor of $\sim 310$ nH. They are time-division multiplexed using superconducting quantum interference device (SQUID) amplifiers made at the National Institute of Standards and Technology (NIST) in Boulder, CO. A series array of SQUIDs at 4 K provides the final amplification of the signal before it is sent to room-temperature electronics. The warm readout Multi-Channel Electronics (MCE)~\cite{2008JLTP..151..908B} that perform the multiplexing have been successfully operated in the field by a number of other experiments.~\cite{2014arXiv1403.4302B, 2009AIPC.1185..494E, 
2011ApJS..194...41S, 2013MNRAS.430.2513H}

\begin{figure}[tbp!]
\centering
  1\includegraphics[width=0.55\textwidth, trim=4in 1.5in 4in 3.25in]{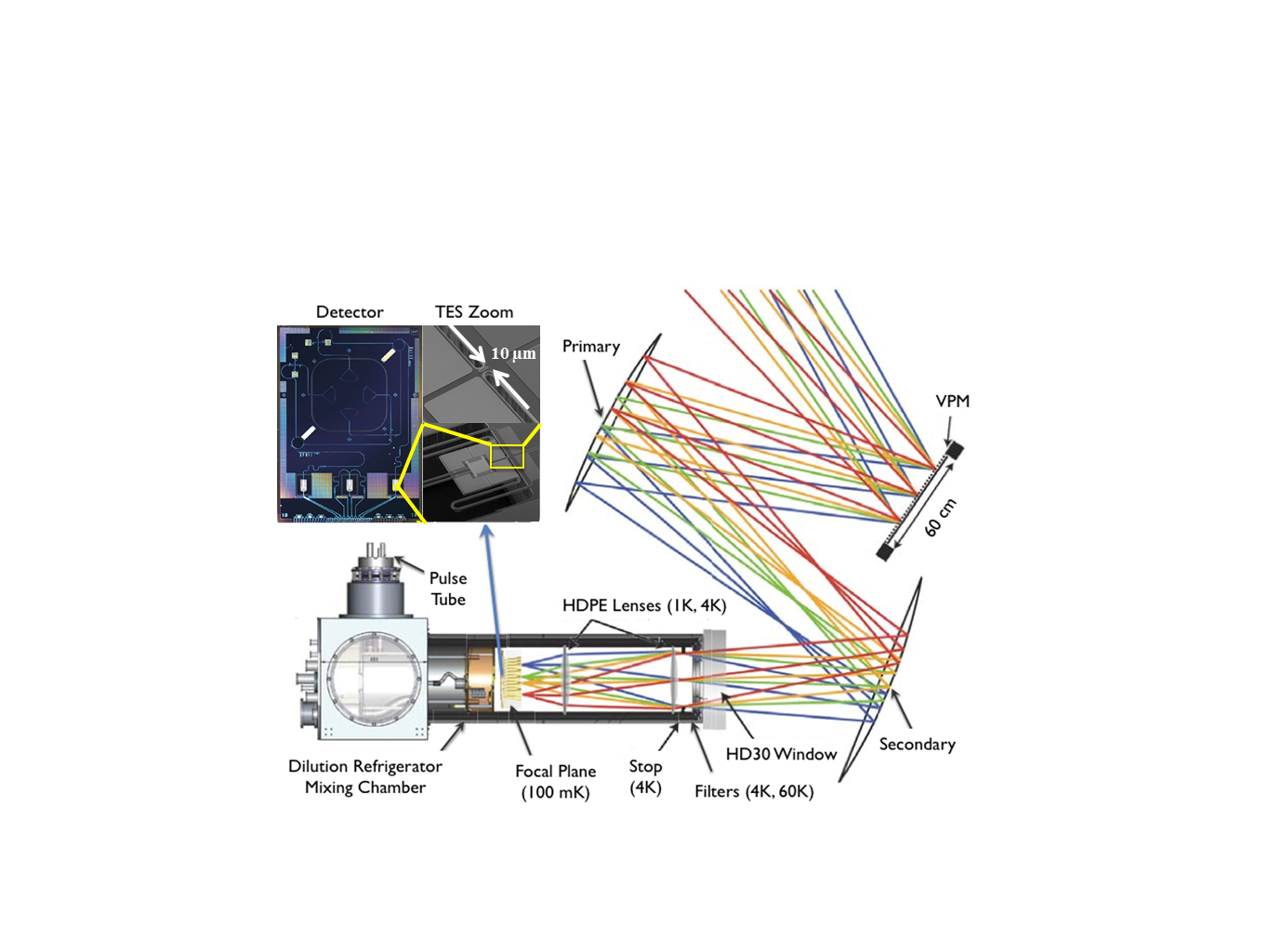}
  \caption{The CLASS telescopes share a common design, with cold refractive optics inside the receiver and warm reflective optics outside. The VPM is the first element that light hits from the sky, modulating celestial polarization but not instrumental polarization. Colored lines show the optical path for four rays converging on four locations in the Q-band telescope. The locations of important optical components and the focal plane are pointed out in the figure. A photograph of a detector and a scanning-electron-microscope image of a Q-band TES are shown in the inset. A zoom-in shows the short, 10$\mu$m-wide bolometer leg that determines the thermal conductance to the TES island.}
\label{fig:class_layout}
\end{figure}

\section{Cryogenic Receivers}
\label{sec:receivers}
The four identical cryogenic receivers for CLASS were made by BlueFors Cryogenics.\footnote{BlueFors Cryogenics, Arinatie 10, 00370 Helsinki, Finland, www.bluefors.com} They are pulse-tube-cooled, dilution refrigerators that provide 60 K, 4 K, 1 K, and 70 mK cold stages, with a 46 cm vacuum-window aperture. This large aperture, combined with the 70 mK temperature of the focal plane, requires careful control of loading from thermal radiation and the development of techniques for producing large-diameter, millimeter-wave optics. The CLASS dilution-refrigerator receivers use a $^{3}$He/$^{4}$He mixture to provide greater than 300 $\mu$W of cooling power for the focal plane at 70 mK, along with 10 mW of cooling power at the 1 K stage. BlueFors custom designed the CLASS cryostats with the ability to be operated horizontally and with long radiation shields to provide a large 4 K volume to allow the second lens in the system to be placed at 4 K. We worked closely with BlueFors to develop their first dilution refrigerator for an astronomical application.

Due to the use of SQUID amplifiers in the detector readout chain the CLASS instrument is sensitive to magnetic fields. Additionally the transition temperature of the TES bolometers is sensitive to magnetic field. The magnetic field of the Earth and magnetic noise from the mount servo motors and other nearby electronics need to be reduced by a factor of at least $\sim100$ to avoid excessive scan-synchronous pickup and noise in the SQUIDs. The receivers incorporate two layers of high-$\mu$ magnetic shielding\footnote{Amumetal 4K, Amuneal Manufacturing Corp., 4737 Darrah St., Philadelphia, PA 19124, www.amuneal.com.}. The first magnetic shield runs most of the length of the 4 K radiation shield. The second shield is placed closely around the focal plane at 100 mK. Together these shields are estimated to provide a factor of 150 shielding at the focal plane. Additional magnetic shielding is provided by a niobium box around the series array and niobium sheets above and below the SQUID multiplexing chips in the focal plane.

\section{Optical Design}\label{sec:optics}
The CLASS telescopes use a combination of warm reflective and cold refractive optics to form beams on the sky with low cross-polarization leakage and minimal warm beam spill.\cite{Eimer2012SPIE} The first optical element in the telescope is the fast polarization modulator, the VPM, ensuring that instrument polarization is not modulated and making CLASS relatively insensitive to possible sources of systematic errors such as lens birefringence. Primary and secondary reflectors, each approximately 1.5 m in diameter, re-image the VPM onto a 4K cold aperture stop inside the receiver. An ultra-high molecular weight (UHMWPE) window provides high throughput in the observing band while holding vacuum in the receiver cryostat. Infrared-blocking filters inside the receiver reject thermal loading from the ambient-temperature surroundings.

The cold aperture stop allows controlled illumination, in the time-reversed sense, of the warm optics, reducing loading from warm beam spill and controlling illumination on the VPM. Two high-density polyethylene (HDPE) lenses, one at 4 K and the second at 1K, focus light onto the focal plane. Smooth-walled copper feedhorns provide single-mode illumination of the detector ortho-mode transducer (OMT), which carries power to the TES bolometers. The layout of the Q-band telescope, with a ray-tracing, is shown in Figure \ref{fig:class_layout}. The W-band telescope will have identical warm optics, with changes to the shape of the HDPE lenses to accommodate a faster feedhorn design. The HF instrument will have silicon lenses with simulated dielectric anti-reflection (AR) layers cut into them.~\cite{2013ApOpt..52.8747D}

\subsection{Feedhorn antennas}
\label{sec:feedhorns}
The CLASS Q- and W- band receivers use novel wide-bandwidth, smooth-walled feedhorns directly machined from oxygen-free high-conductivity (OFHC) copper. The smooth-walled design significantly reduces the complexity and cost of producing the feeds, which is particularly important for the W-band receivers where a total of 518 individually-machined feedhorns are required. The design of the feedhorns was optimized using custom code that solved the boundary conditions for EM waves in concentric cylindrical waveguide sections, with a TE$_{11}$ circular waveguide mode input at the base of the feedhorn.~\cite{2010ITAP...58.1383Z}

\begin{figure}[t!]
\centering
\subfloat[]{
		\includegraphics[height=0.35\textwidth, clip=true, trim=0in 0.1in 0in 0.25in]{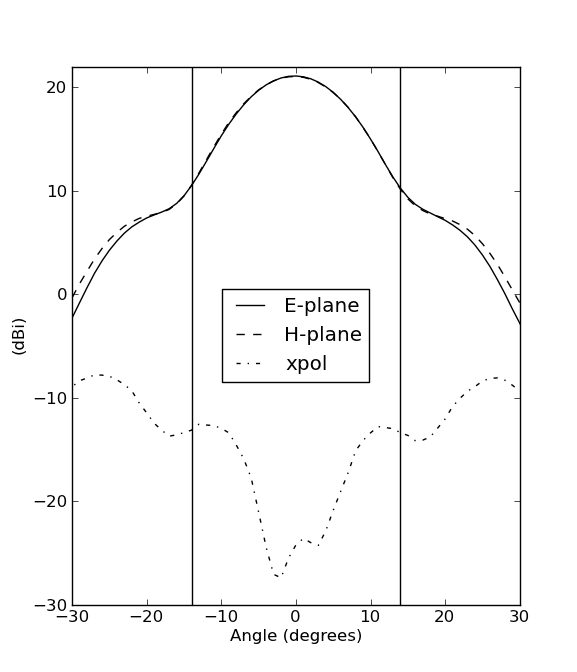}
\label{fig:qband_feed}
}
\subfloat[]{
		\includegraphics[clip=true, trim=0 0 0 0, angle=0, height=0.33\textwidth]{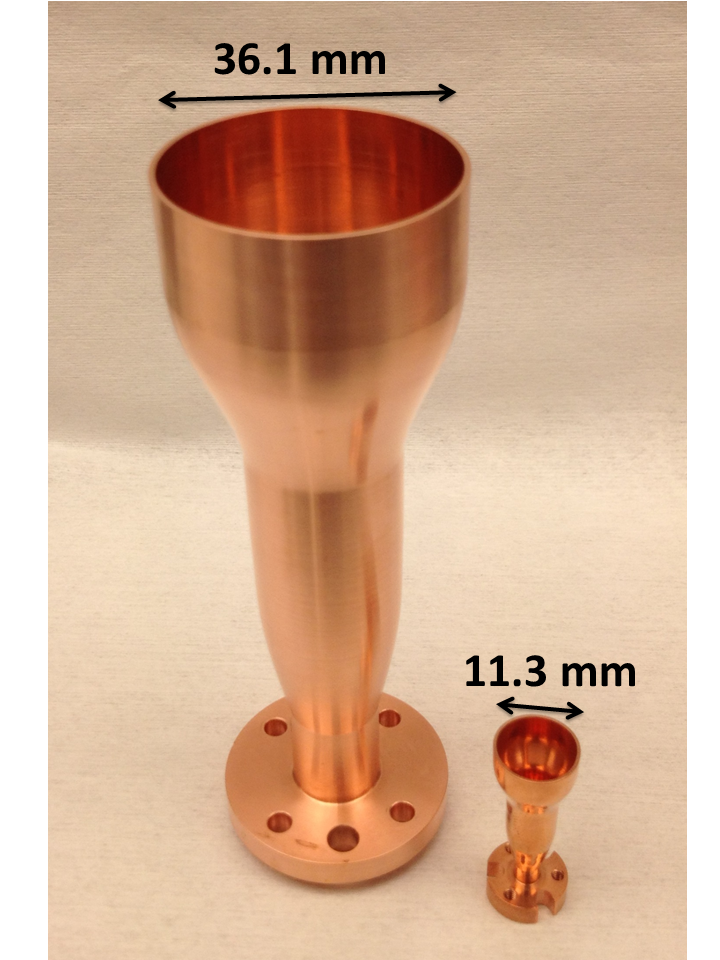}
		\label{fig:feed_pic}}
\subfloat[]{
		\includegraphics[clip=true, trim=0 0.1 0 0.25in, angle=0, height=0.35\textwidth]{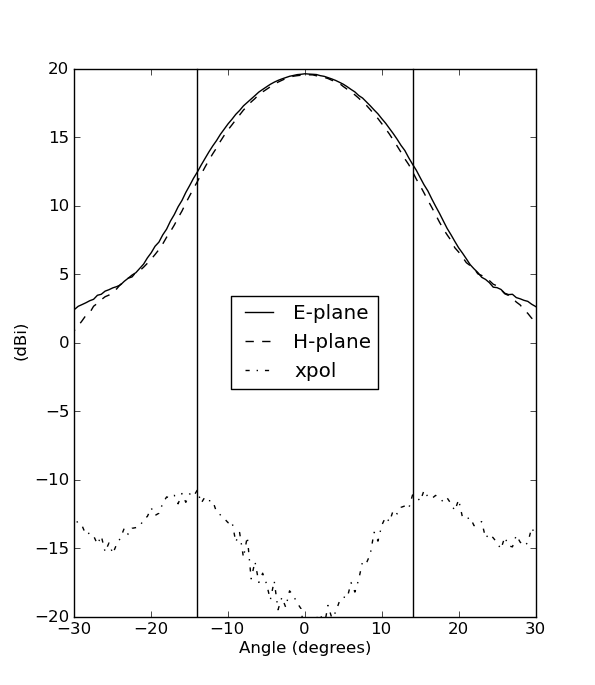}
		\label{fig:wband_feed}}
\caption{(a) Q-band feedhorn copolar E- and H-plane and crosspolar beams averaged across the 33-43 GHz band, as measured in the Goddard Electromagnetic Anechoic Chamber (GEMAC). Vertical lines are at $\pm 14^{\circ}$ where the beams truncate on the receiver cold stop. Measured and modeled beams agree to within $\pm$2\% in this region. The Q-band feed has a beam wtih FWHM of $14.0^{\circ}$ (b) Photograph of Q- and W-band feedhorns. (c) Band-averaged beams measured for the W-band prototype feedhorn. As with the Q-band feed, the measured and modeled beams agree to within 2\%. The W-band feed has a beam with $18.7^{\circ}$ FWHM.}
\label{fig:feeds}
\end{figure}

Prototypes were made of the Q and W feedhorn designs and the beam pattern was measured in the Goddard Electromagnetic Anechoic Chamber (GEMAC). The measured beams matched the model to within 2\% over the region $\pm 14^{\circ}$ not truncated by the cold aperture stop, for both cases. Figure~\ref{fig:feeds} shows the measured and modeled beams for the two feedhorns. The Q-band feeds have a band-averaged 14.0$^{\circ}$ beam FWHM. The W-band feeds were designed for closer packing to accommodate more detectors per 100 mm wafer, with the result that the feedhorn has a wider, 18.7$^{\circ}$ FWHM, beam than at Q band. The Q-band feeds have all been made and are ready for installation in the focal plane. The W-band feeds are in production.

\subsection{Cryogenic lenses}
CLASS employs cryogenic lenses in all its frequency channels. HDPE was chosen for the Q- and W-band CLASS lenses because it is easily machined, has low millimeter-wave loss, and is readily anti-reflection coated. Literature values~\cite{1981InfPh..21..225B, 1991ITMTT..39..352S} for HDPE give its index of refraction as $n=1.53$ at room temperature. HDPE contracts by approximately 2\% when cooled to 1 or 4 K, which leads to a shift in the index to an estimated $n = 1.56$. At approximately 5 cm thick, each lens attenuates $\sim$1\% over the Q band and $\sim$3\% over the W band. The HF instrument will have lenses made of silicon with AR layers cut into them with a dicing saw,~\cite{2013ApOpt..52.8747D} as the loss through thick HDPE becomes prohibitive at the higher frequencies. Due to its high index of refraction, silicon lenses can be made thinner for lower in-band loss. The first lens is mounted from the 1 K stage of the cryogenic receiver, while the second lens is mounted on the 4 K stage. Because they are cold, the HDPE lenses add negligible additional optical loading to the detectors.

The HDPE lenses for CLASS are machined on a CNC mill at JHU. The same mill is used to machine a simulated-dielectric AR layer. The AR layer is an array of cylindrical holes machined into the surface of the lens in a square grid to create an effective dielectric with an index of refraction $n \sim 1.25$ that is $\lambda/4n$ thick. The Q-band AR layers are made of holes 1.54 mm in diameter drilled 1.63 mm deep on a grid spacing of 1.836 mm. This is modeled to give 0.16\% band-averaged reflection. The two Q-band lenses will have approximately 0.7\% band-averaged absorption. The W-band AR layers are made of holes 0.81 mm in diameter drilled 0.67 mm deep with a grid spacing of 0.96 mm. This is modeled to give band-averaged reflection of 0.2\% and absorption of 1.6\% for both lenses combined. The depth of the holes does not correspond to exactly $\lambda/(4 n)$ because the drill bits have a tapered end. This has been modeled as a two-layer AR coating with the hole depth as a free parameter.

In order to relieve stresses in the HDPE that could deform the lenses, the HDPE lenses will be annealed, then rough-cut, then annealed a second time before the final lens surfaces are cut. During annealing, the material temperature is raised to 120$^{\circ}$C over the course of 4 hours, left there for 48 hours, and then cooled to room temperature over 48 hours. 

\begin{figure}[t]
\centering
\subfloat[]{
		\includegraphics[height=0.38\textwidth, clip=true, trim=0in 0in 0in 0in]{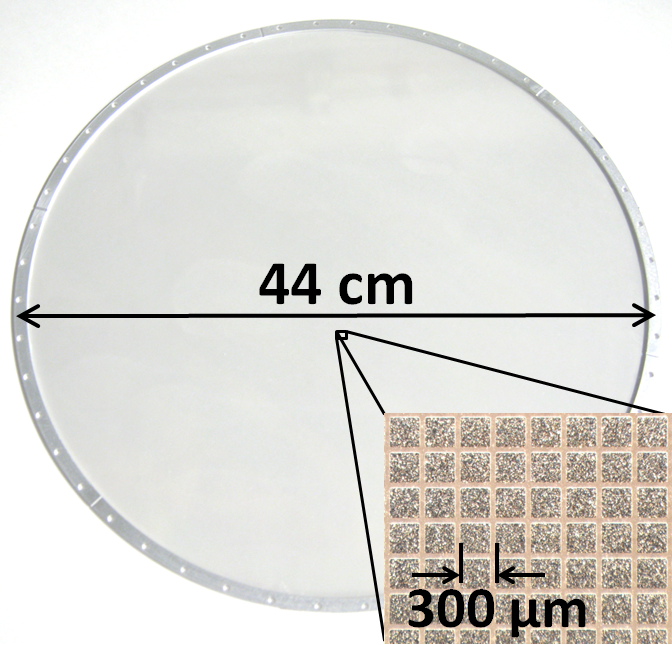}
\label{fig:mmf_filter}
}
\subfloat[]{
		\includegraphics[clip=true, trim=0 0 0 0, angle=0, height=0.38\textwidth]{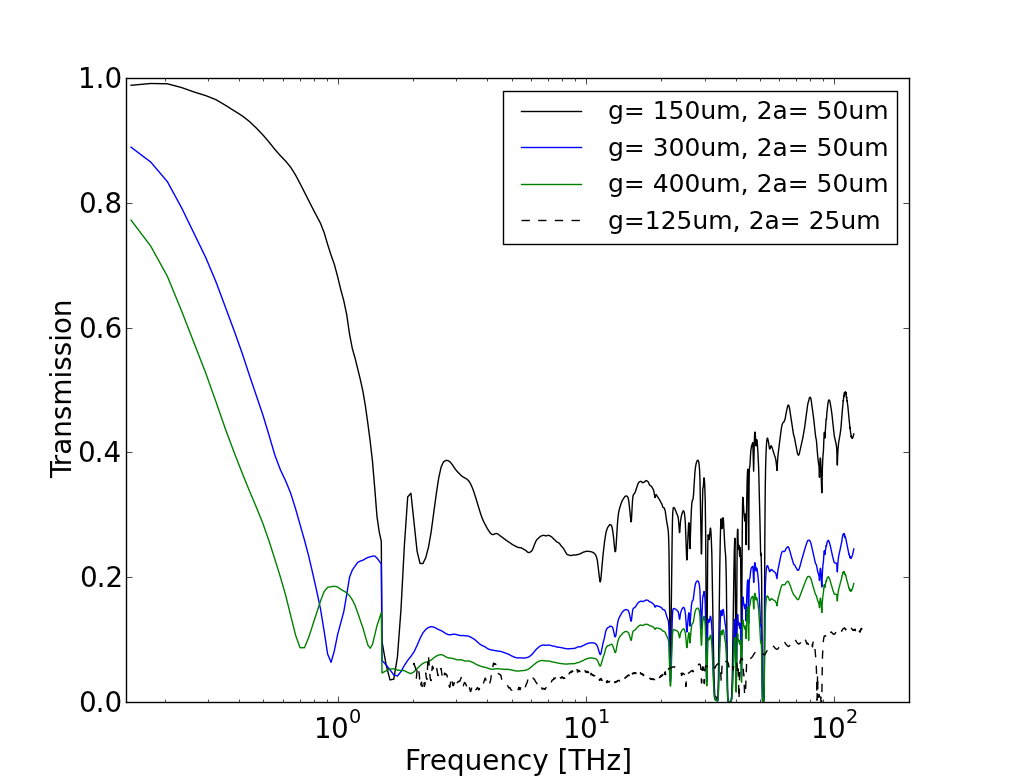}
		\label{fig:mmf_spectra}}
\caption{(a) A photograph of a class MMF on its mounting ring. The aperture is 44 cm. The inset shows a photograph taken with a microscope of the grid pattern, which in this case has a grid spacing of 300~$\mu$m. (b) Measured spectra of CLASS MMF, taken with a combination of a terahertz time-domain spectrometer~\cite{2012arXiv1203.5019M} for frequencies below 1 THz and an infrared Fourier-transform spectrometer for frequencies above 1 THz. The spectra for four grid patterns are shown. Solid lines indicate patterns on 6 $\mu$m mylar made photolithographically, while the $g$$=$125 $\mu$m pattern is on 14.6 $\mu$m polypropylene (PP) and was made with laser delamination. The frequency cutoff for these lowpass filters occurs at a wavelength $\lambda \sim 1/g$. A clear trend is seen at high frequency for transmission to go as approximately the metal filling-fraction of the pattern, $f=g^2-b^2$, where $b$ is the square side length. The strong absorption lines of mylar near the peak of the 290 K blackbody at 30 Thz are visible in the solid curves. This strong absorption makes mylar a less-attractive substrate material compared to PP.}
\label{fig:filters}
\end{figure}

\subsection{Infrared-blocking filters}
Nearly 40 W of infrared power is expected to be incident on the receiver vacuum window from the surrounding ambient-temperature enclosure. This power needs to be reduced to no more than 100 $\mu$W by the 70 mK stage where the focal plane resides. This means the IR power needs to be cut by a factor of 400,000, while negligibly reducing in-band power. The maximum power that can be dissipated on the ends of the 4 K and 60 K radiation shields is 0.5 W and 9 W, respectively. This requires that 75\% of incident IR power be reflected out of the receiver cryostat immediately, while the remainder can be absorbed and conducted out of the 60 K radiation shield. Only a small fraction can be conducted away by the 4 K radiation shield. 

Metal-mesh filters (MMF) are used by CLASS to reflect away IR power. The CLASS MMF were produced in part by photolithography of 6 $\mu$m mylar film coated with 300-500 \AA~of aluminum.\footnote{Performed by Tech-Etch, Inc., 45 Aldrin Road, Plymouth, MA 02360 USA, www.tech-etch.com}. The remainder of the MMF are fabricated by laser delamination of aluminized 12.6 $\mu$m polypropylene (PP) film.\footnote{Performed by Photomachining, Inc., 4 Industrial Dr., Unit 40, Pelham, NH 03076 USA, www.photomachining.com } The use of PP instead of mylar reduces in-band loading, especially for the HF instrument. Additionally, the laser delamination process can produce cut line widths in the aluminum as narrow as 25 $\mu$m, compared to 40 $\mu$m for the photolithographically-produced filters. As can be clearly seen in Figure~\ref{fig:mmf_spectra}, this translates into lower transmission at high frequencies, where the capacitive grid no longer resonates, but rather transmits approximately in proportion to empty area in the grid pattern. HFSS\footnote{Ansys, Inc., www.ansys.com } modeling and spectrometer measurements bare out this conclusion. The in-band and IR transmission of the metal-mesh filters for CLASS have been validated in spectrometers at JHU. 

The spectrum of a selection of CLASS MMF are shown in Figure \ref{fig:mmf_spectra}. In that figure, one can see common characteristics of these filters. At low frequencies, the square elements in the capacitive grid are too small compared with a wavelength to efficiently couple to the incoming radiation. This ensures that the MMF are highly transmissive at millimeter wavelengths. As the wavelength approaches the size of the grid elements, transmission falls steeply until a resonance is reached in which nearly all radiation is reflected. At high enough frequencies the grids transmit roughly in proportion to the empty area in the grid pattern.

After two MMF reflect away a majority of power at the vacuum window itself, a filter stack at 60 K reduces the remaining power below the 0.5 W allowed to transmit to the 4 K stage of the receiver. This 60 K filter stack consists of two PTFE filters, each 12 mm thick and AR coated using a simulated-dielectric technique like that used for the lenses. Above, below, and in between the PTFE filters are a pair of MMF. A filter testbed has been made to validate the 60 K filter design. This testbed consists of a 40-cm aluminum plate with absorber tiles epoxied onto the top of it. The plate is then thermally isolated from a liquid-nitrogen (LN2) cooled baseplate by thin Z-shaped supports made of stainless steel. This provides a weak thermal link to the LN2 bath. A diode thermometer reads the temperature of the baseplate, which is proportional to power incident on the absorber tiles. The filter stack is placed above the absorber, blocking the 300K radiation of the cavity around it. With this testbed, the 60 K filter stack for CLASS has been shown to allow no more than 0.5 W of power through. This is an upper limit on the power transmitted through the filter stack, as the LN2 bath only cools the edge of the filter stack to 85 K, instead of the 60 K provided by the receiver cryostat.

Additional thermal filters will be placed at the 4 K stage to reduce power to the 1 K and 70 mK stages. This will consist of one additional 12-cm PTFE filter and four MMF. A final nylon filter at 1 K will provide thermal filtering below 1 THz if necessary. This may be required because PTFE does not efficiently absorb below 1 THz. 

\subsection{Vacuum window}
The vacuum window for CLASS needs to provide high in-band transmission while holding vacuum over the 46-cm aperture of the CLASS receivers. The CLASS windows are made of 4.8-mm-thick, ultra-high molecular weight polyethylene (UHMWPE) and will be AR coated with porous PTFE. The PTFE will be heat pressed onto the UHMWPE using a thin film of low-density polyethylene (LDPE). A two-layer AR coating is used to provide wide-bandwidth performance for the HF instrument. A prototype window has passed vacuum tests and was shown to bow approximately 6.4 cm at atmospheric pressure. The pressure at the high-altitude site is approximately half of standard atmospheric pressure, meaning the window should bow less under observing conditions. Monitoring of the window for an extended period showed that creep in the material stopped after approximately one week. The UHMWPE window for CLASS is expected to introduce 0.2\% absorption and 0.15\% reflection averaged across the Q band, and 0.45\% absorption and 0.2\% reflection averaged across the W band.

\subsection{Primary and secondary reflectors}
The 1.5-meter primary and secondary reflectors for the CLASS Q and W telescopes have been machined out of monolithic slabs of aluminum using a large CNC mill. Figure~\ref{fig:reflector_pic} shows a photograph of a completed secondary reflector. The reflectors are $\sim 10$ cm thick, with significant light-weighting of the back, leaving a series of 5-mm-thick ribs for stiffness, leaving the reflector aluminum 3 mm thick in between the ribs. Both mirrors are off-axis ellipsoids. For the exact geometry, refer to a previous publication.~\cite{Eimer2012SPIE}

The reflectors are oversized for the resolution of the telescope to minimize warm beam spill, reducing loading on the detectors and sources of instrumental polarization from uncontrolled surfaces. This keeps the estimated warm spill below 0.5\% and 0.4\% for the Q and W telescopes, respectively. A set of three leaf springs are mounted on the back of each reflector and connect the reflector to the telescope structure while allowing adjustment of the reflectors.

\subsection{Co-moving ground shield}
As shown in Figure~\ref{fig:class_mockup}, the receiver and the warm, reflective optics are enclosed in a co-moving ground shield. This (1) ensures that warm beam spill falls on stable surfaces, greatly reducing ground pickup and far-sidelobe response to sources of strong emission, particularly the sun; (2) provides a clean environment for the VPM and other optics, keeping dust and snow from accumulating on them; and (3) allows for temperature control of the warm optics as a means of reducing $1/f$ noise in the detector timestreams.

The co-moving ground shield is made of a frame of aluminum beams with a skin of aluminum honeycomb panels bolted onto it. The panels are sealed at their edges. The inside of the ground shield and the warm optics is thermally regulated to reduce $1/f$ noise from thermal drift in the optics and provide a stable temperature for any warm beam spill.

\begin{figure}[t]
\centering
\subfloat[]{
		\includegraphics[height=0.45\textwidth, clip=true, trim=12in 3in 22.5in 0in]{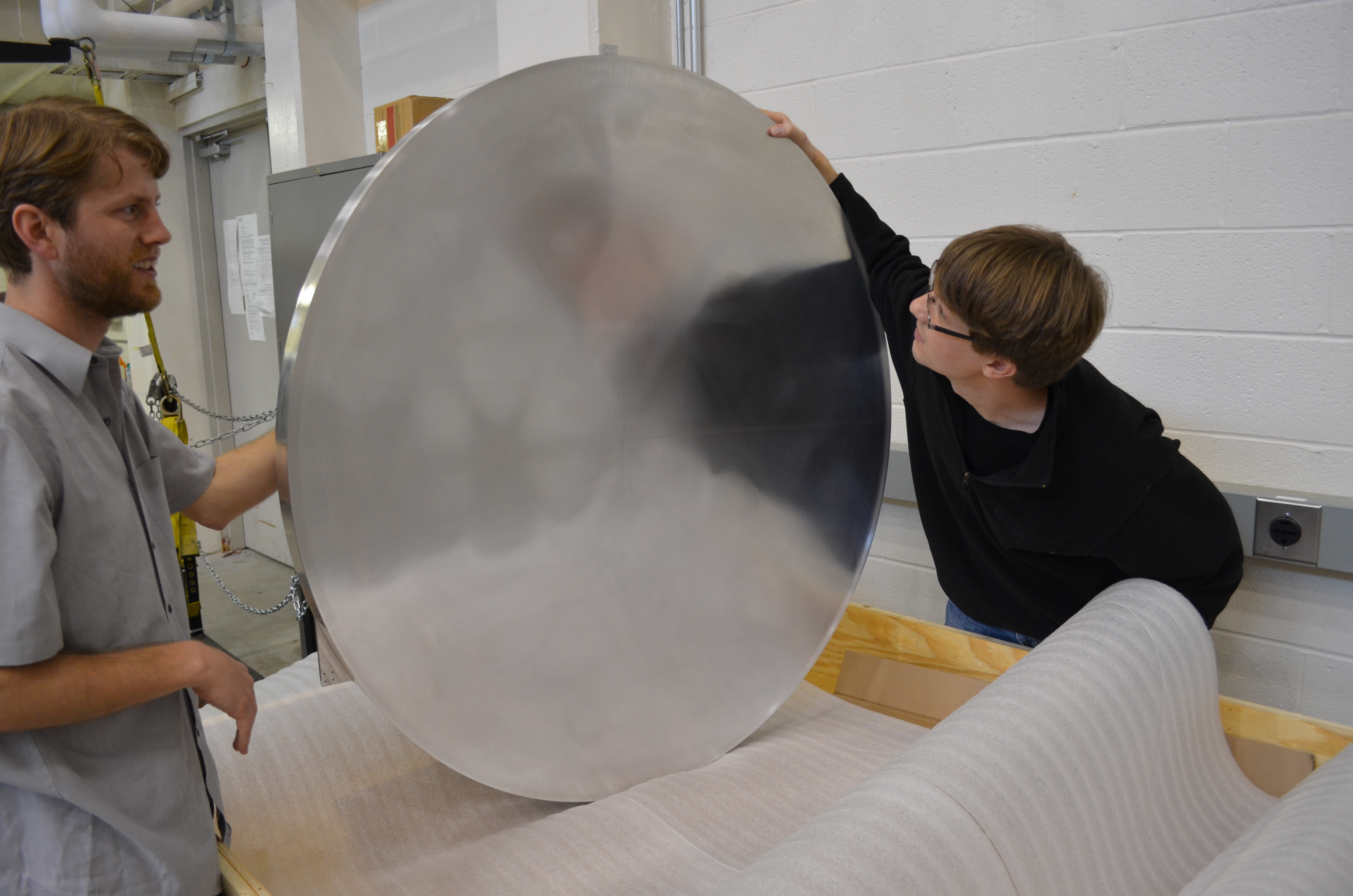}
\label{fig:reflector_pic.jpg}
}
\subfloat[]{
		\includegraphics[clip=true, trim=0 0 0 0, angle=0,height=0.45\textwidth]{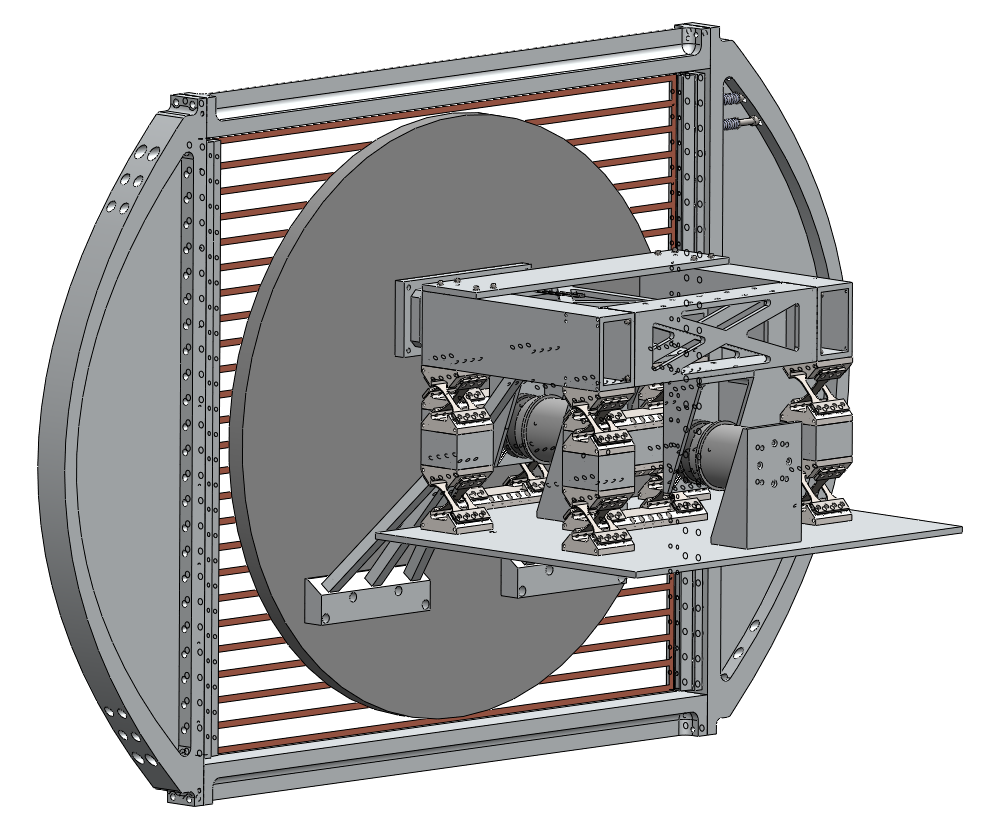}
		\label{fig:vpm_solidworks}}
\caption{(a) Photograph of a secondary reflector. (b) The mechanical design for the VPM consists of a resonant four-bar linkage flexure attached to the mirror. A reaction-canceling mass will be moved in opposition to the mirror to reduce vibrations that could feed into detector timestreams. }
\label{fig:vpm_mechanical}
\end{figure}

\section{Variable-Delay Polarization Modulators (VPMs)}
\label{sec:vpm}
The use of fast polarization modulation is crucial in allowing CLASS to recover large-angular-scale modes on the sky, because it separates the polarized signal of interest from the total sky brightness and the associated $1/f$ noise from the atmosphere. Such fast modulation was an integral component of coherent detector arrays~\cite{2003ApJS..145..413J, 2004ApJ...610..625F, 2005ApJS..159....1B, 2012ApJ...760..145Q, 2010A&A...520A...4B, 2006PhDT........33S, 2003PhRvD..68d2002O, 2005ApJ...624...10L, CBI_instrument_2002, 2009ApJ...694.1664C, 2013ApJ...765...64M} and has been demonstrated with good success on other bolometric instruments, notably with the Atacama B-Mode Search (ABS)~\cite{2014RScI...85b4501K}, which employs a warm, continuously-rotating half-wave plate (HWP) as a modulator.

\begin{figure}[t]
\centering
\subfloat[]{
		\includegraphics[height=0.38\textwidth, clip=true, trim=0.2in 0.2in 0.2in 0.2in]{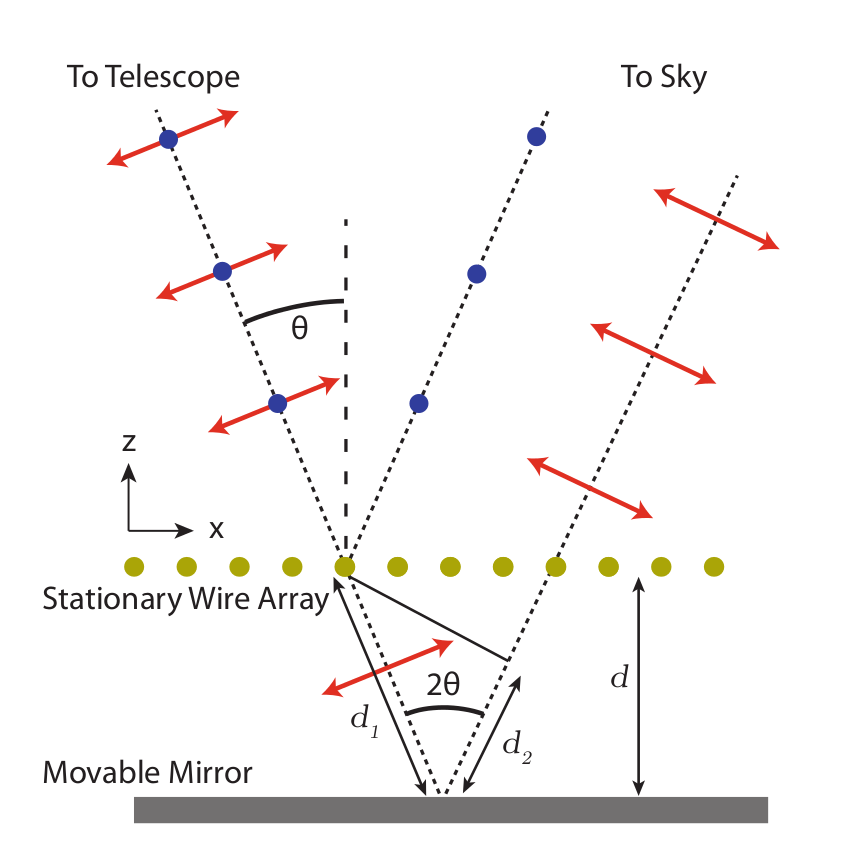}
}\subfloat[]{
		\includegraphics[height=0.38\textwidth, clip=true, trim=0in 0in 0in 0.5in]{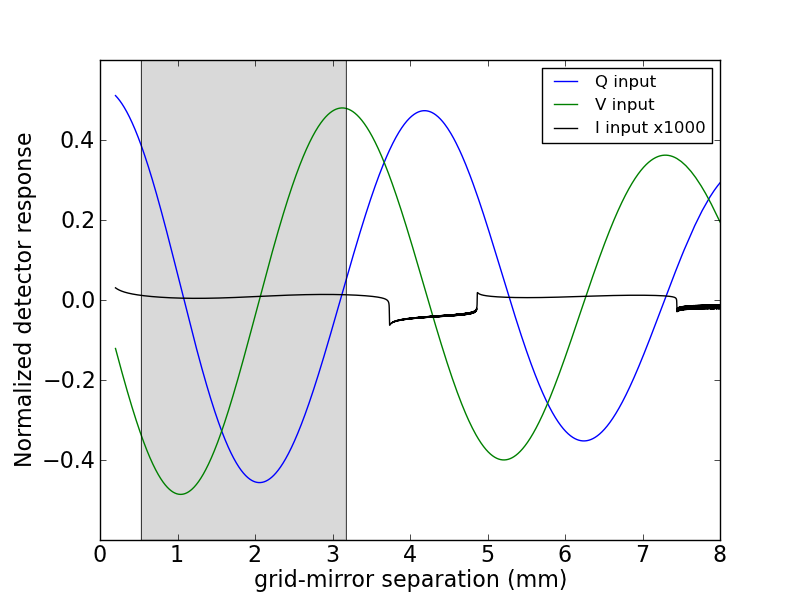}
		\label{fig:classq_transfer_funcs}}
\caption{(a) The VPM consists of a stationary wire grid (gold dots) with a movable flat mirror. The component of the incoming signal that is polarized parallel to the grid wires is reflected by them. The orthogonal linear polarization passes through the grid and is reflected by the mirror. This introduces a phase difference $\delta = 2 k d \cos\theta $ between the two orthogonal linear polarizations and modulates incoming linear to circular polarization, and vice versa. (b) Band-averaged modeled detector response at Q band versus grid-mirror separation for unity-amplitude input unpolarized (Stokes I), linearly polarized (Stokes Q), and circularly polarized (Stokes V). Stokes U is not modulated. All transfer functions have had their mean removed. The transfer function for Stokes I is multiplied by 1000 to make it visible on the same scale and by Kirchoff's Law gives the estimated modulated polarized grid emission from the VPM.}
\label{fig:odd_combo}
\end{figure}

The variable-delay polarization modulator (VPM)~\cite{2012ApOpt..51..197C, 2006ApOpt..45.5107C, 2014arXiv1403.1652C} technology that will be used by CLASS differs from other modulation techniques in a few key ways. It uses only reflective optical elements, allowing its use at ambient temperature without the sensitivity hit associated with a warm HWP or other dielectric element. This greatly facilitates its placement as the first optical element in the system, which is desirable for the separation of celestial from instrumental polarization. Additionally, it modulates Q to V Stokes parameters, and vice versa. Because there is not expected to be V polarization in the CMB or from the Galaxy at large angular scales, the VPM provides a valuable null channel for systematics checks.

Each CLASS VPM, see Figure~\ref{fig:vpm_solidworks}, consists of a wire array backed by a lightweight aluminum honeycomb mirror. During modulation, the wire array remains fixed and the mirror translates towards and away from the array sinusoidally at $10$ Hz while remaining parallel to it. The amplitude of the modulation is optimized for each CLASS band with the maximum throw of $1.5$ mm for the 40 GHz band. 

Parallel transport of the mirror is achieved by a four-bar-linkage flexure attached to the back surface of the mirror. Each of the rotary joints of the flexure consist of wire-EDM machined crossed straps. Blue-tempered spring steel was selected as the strap material to decrease the risk of fatigue failure over the $\sim10^9$ cycles expected during the life of the CLASS survey. Each of the straps is located by pins and bolted into a stainless steel fixture to precisely locate the rotation axis of the rotary joint. The natural frequency of the mirror-flexure system is tuned to be near the 10 Hz modulation frequency. 

An analogous flexure system will operate $90^\circ$ out of phase of the primary flexure to cancel the reactive forces and minimize vibrational coupling to the telescope. Each flexure is actuated by a linear voice coil.\footnote{ BEI Kimco Magnetics LA25-42-000A} Closed-loop control of the modulation is achieved by measuring both the mirror position and the reaction canceling flexure position with linear encoders\footnote{Renishaw ATOM Ri 0200 } and fed back through a control system to precisely maintain sinusoidal movement. A set of three of the same linear encoders are positioned around the rear face of the mirror to monitor residual tip and tilt motion, and they are read out synchronously with the detectors at 200 Hz for subsequent demodulation of the polarization signal.

The wire grid itself consists of nearly 3900 copper plated tungsten wires, 50.8 $\mu$m in diameter, regularly spaced at 159.5 $\mu$m. The ends of the wires are fixed with Stycast 2850 to bars running the width of the wire array. Uniform wire spacing is ensured by placing each wire into a groove cut into these bars. This fully-strung wire array is tensioned via coil springs between support bars with an I-beam cross-section. Each wire is tensioned to $\sim$2.2 N, $30\%$ of its breaking strength, delivering a total of 8470 N load to the support bars. To limit overall deflection, the outer portion of the support beam is catenary shaped with both the inner and outer arching surfaces specified as a compromise between packaging the VPM within the support structure and minimizing deflection under the load. The plane of the wires is defined by a separate, optically-flat, 60 cm clear aperture ring gently pressed against the wire array. Micrometer adjusters allow the plane of the wires, defined by the flattening ring, to be aligned parallel to the movable mirror.

The wire array, flattening ring, and flexure assembly are each kinematically mounted to a stiff aluminum weldment frame. The entire VPM assembly mounts through three pivot joints on this weldment frame to the rest of the telescope structure allowing for adjustment of the final direction of the telescope pointing. Lightweight paneling encloses the VPM structure - opening only on the mirror side of the VPM - and the temperature of the system is controlled to ensure stable performance of the electronics and the encoders while simultaneously stabilizing the transfer function of the VPM. 

This transfer function for the VPM, averaged over the 33-43 GHz Q band, for input unpolarized, linearly-polarized, and circularly-polarized radiation with unity power is shown in Figure~\ref{fig:classq_transfer_funcs}. Q and V are modulated at high efficiency out of phase with one another, allowing easy separation of linearly- and circulary-polarized light. The sinusoidal waveform of the VPM mirror motion goes between the edges of the gray region at approximately 10 Hz. Because the mirror motion spends more time proportionally at the ends of its range of motion, which are chosen to be points of maximum Q signal, the VPM modulation efficiency to Q is 85\%, instead of the 50\% one would get with equal time spent on Q and V. Note that the unpolarized (black) transfer function has been multiplied by a factor of 1000 to make it visible on the same scale as the Q and V input transfer functions. Input U, corresponding to polarization at $\pm 45^{\circ}$ to the wire grid, is not modulated by the VPM.

\section{Mounts and Atacama Site}
\label{sec:mount_and_site}
CLASS will employ two altazimuth mounts with a third, boresight, axis of rotation to allow regular rotation of the polarization direction of the detectors relative to the sky. The mounts are being outfitted and tested in a high bay at JHU, where full system integration will take place before the telescope is shipped to Chile. The mounts have 1800 lb capacity and are capable of 5$^{\circ}$/s slew speeds maintaining 2$^{\prime}$ pointing accuracy under continuous winds of 13 m/s gusting to 20 m/s. A platform that rotates with azimuth houses the gas-handling systems for the receivers, mount servo system electronics and power distribution, pulse-tube helium compressors, and housekeeping electronics. 

Site construction will commence later this year, including installation of the pedestal foundations, generator buildings, radio link to the neighboring town of San Pedro de Atacama, and wiring conduit. The CLASS site is near the Atacama B-Mode Search (ABS), Atacama Cosmology Telescope (ACT), and Polarbear sites on Cerro Toco at an altitude of approximately 5140 m. Power for the site will be provided by diesel generators. The site will have work areas and laboratories made from shipping containers, including a receiver-assembly lab, a control and operations container, and a machine shop.

\begin{figure}[t]
\centering
\subfloat[]{
		\includegraphics[height=0.35\textwidth, clip=true, trim=0.2in 0.2in 0.2in 0.2in]{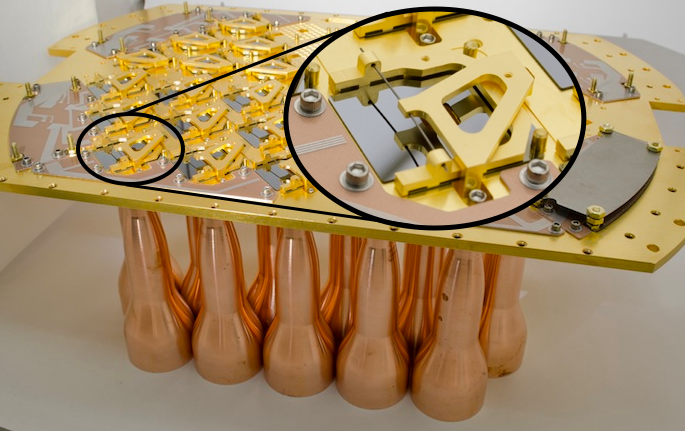}
\label{fig:q_focal_plane.png}}
\subfloat[]{
		\includegraphics[clip=true, trim=1.5in 0 0.5in 0, angle=0, height=0.35\textwidth]{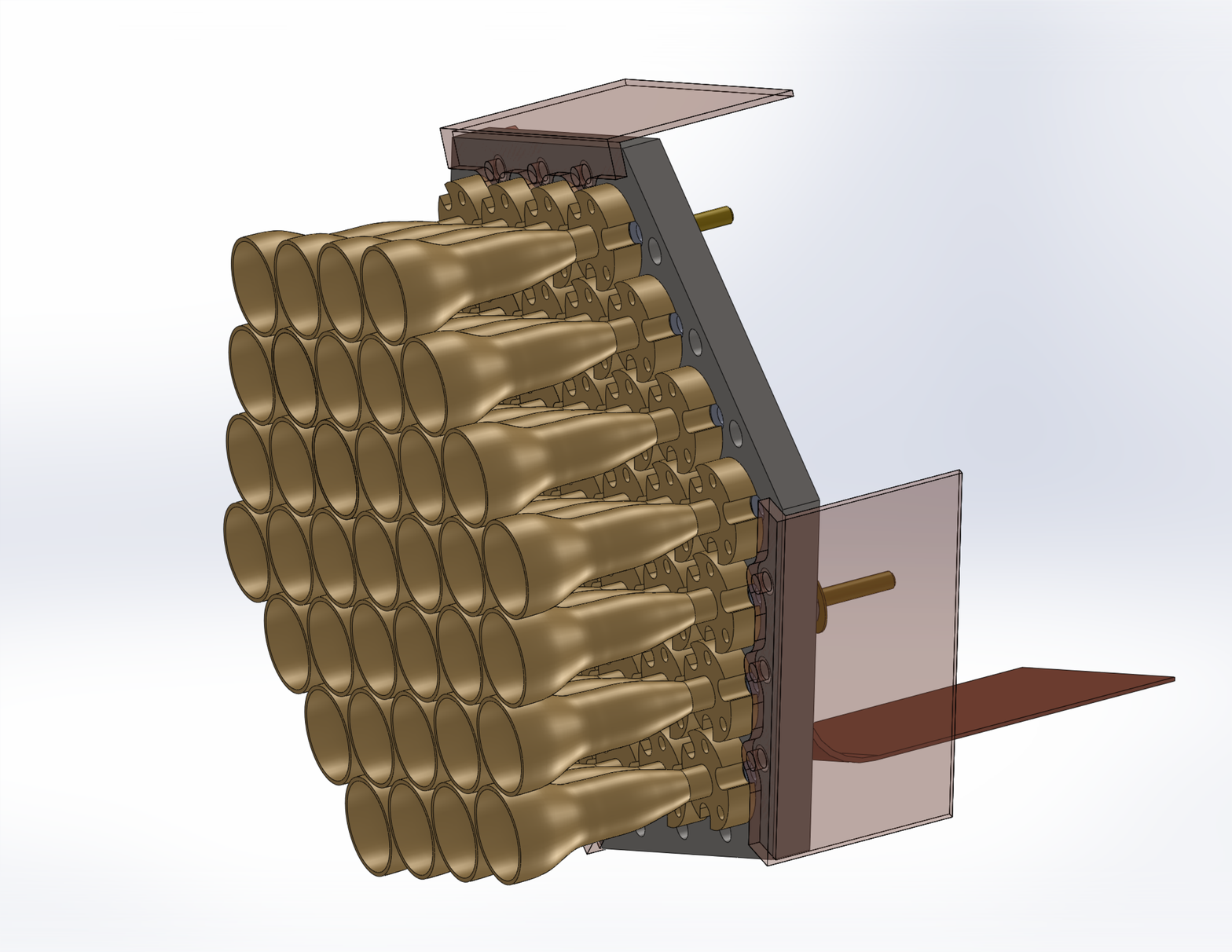}
		\label{fig:w_wafer_module}}
\caption{(a) Photograph of the Q-band focal plane partly assembled. The zoom-in shows an individual detector clip holding down a dummy wafer. All detector signals are carried to the edges of the baseplate where the four multiplexing (MUX) SQUID electronics boards reside. Each MUX circuit is magnetically shielded with niobium sheets above and below. A subset of the feedhorns were installed for this photograph. (b) Concept drawing of the W-band monolithic wafer module with 37 detectors pairs. The feedhorns for the two W-band focal planes are currently in fabrication. The baseplate (in gray) that interfaces between the feedhorns and the detector wafer is planned to be made of a silicon-aluminum alloy. Seven such modules will comprise a focal plane of 259 detectors.}
\label{fig:focal_planes}
\end{figure}

\section{Q-band receiver}
\label{sec:qband_receiver}
The assembly and testing of the Q-band (40 GHz) receiver are well under way, with deployment to the Atacama Desert of Chile expected in early 2015. Many components are already fabricated and validated, while others are in advanced stages of design. Most of these elements were discussed in previous sections. The system was designed for high efficiency between the detector and the sky. It is expected that 68\% of the photons incident on the detector come from the sky. Keeping this system efficiency, $\alpha_{s}$, high is important, because array sensitivity scales as $\alpha_{s} \sqrt{N_{det}}$. Thus, one gains array sensitivity faster by increasing individual detector efficiency than by increasing the number of detectors; the Q-band CLASS telescope can be quite sensitive with just 36 feedhorn-coupled bolometer pairs. The high system efficiency of the Q-band telescope is achieved through the use of a cold aperture stop at 4 K that controls the beam spill on warmer elements, which is additionally helped by the use of over-sized primary and secondary reflectors. Table~\ref{tbl:qband_loading} shows expected efficiencies and beam spill percentages on an element-by-element basis for the Q-band telescope. See a previous paper for details of the Q-band optical design.~\cite{Eimer2012SPIE}

The unique element of the Q-band telescope compared to other frequency channels is the focal-plane design, which is one of the few aspects of the CLASS design that is not shared between frequency channels. A photograph of the Q-band focal plane is shown in Figure~\ref{fig:q_focal_plane}. For this channel, 36 individual polarization-sensitive detector chips are mounted onto a monolithic gold-plated copper baseplate. This baseplate provides an interface between the feedhorns and the detector ortho-mode transducer (OMT), providing a transition from the circular waveguide at the base of the feedhorn and the square waveguide at the OMT. There is additionally a photonic-choke structure machined on the detector side of the baseplate, which combined with a metalized ``choke chip'' under the detector chip minimizes signal loss between the baseplate waveguide and the detector OMT.\cite{wollack_photonic_choke_2010} A novel spring-clip design presses the detector chip against the photonic choke of the baseplate and registers the detector OMT to the baseplate waveguide using a titanium (Ti-6Al-4V) wire as the spring element. More details are given elsewhere in these proceedings.~\cite{appel_spie_2014}

\section{W-band Receiver}
\label{sec:wband_receiver}
The W-band telescopes are similar to their Q-band counterparts with the exception of the focal plane and the details of the anti-reflection coating of the refractive optics and filters. The W-band focal plane architecture for CLASS employs baseplates made of an Al-Si alloy\footnote{Sandvik Osprey, controlled-expansion CE7 alloy.} with a coefficient of thermal expansion (CTE) that is a closer match with that of the detector silicon than a copper or aluminum baseplate would be. This enables the $\sim$100mm silicon wafers with 37 W-band detectors to be mounted directly to the Al-Si with simple spring-loaded clips. The 37-element wafers for CLASS will be assembled into interchangeable modules. Seven modules will be assembled into the focal plane for each of the two 90 GHz receivers, for a total of 259 detector pairs per receiver. Modules to test the fully integrated W-band detector wafers are currently in fabrication. Smooth-walled copper feeds for W-band are also being fabricated. As described in Section~\ref{sec:feedhorns}, a prototype W-band feed had its beam profile measured and was found to be within specifications. Some details of the W-band monolithic detector wafer design is presented elsewhere in these proceedings.~\cite{rostem_spie_2014}

\section{Conclusion}
\label{sec:conclusion}
We have described the Cosmology Large Angular Scale Surveyor (CLASS), a novel ground-based experiment for detecting the signature of inflationary gravitational waves through their impact on the polarization of the CMB. CLASS will map 70\% of the sky in four frequency bands centered on 38, 93, 148, and 217 GHz that straddle the Galactic foreground minimum and the CMB maximum. CLASS will be able to constrain the tensor-to-scalar ratio at a level of $r=0.01$ in the presence of polarized Galactic foreground emission. This is achieved through a system design that maximizes per-pixel detector sensitivity and is optimized for control of systematic errors. The ability of CLASS to make high-fidelity measurements of CMB polarization at the largest angular scales from a ground-based platform is enabled by the use of 10 Hz polarization modulation using variable-delay polarization modulators (VPMs). End-to-end simulations demonstrate our ability to recover Stokes Q and U at low $\ell$ in the presence of a variety of systematic errors and sources of $1/f$ noise.

CLASS will search for $B$-mode polarization in the CMB over the entire range of angular scales of interest for detecting inflationary gravitational waves, from the reionization bump at $\ell < 10$ to the recombination peak at $\ell \sim 80$. This is essential for verification of any primordial $B$-mode detection, and coupled with the ability of CLASS to discriminate CMB from Galactic foregrounds, will allow CLASS to put tight constraints on $r$. Additionally, CLASS will make a cosmic-variance-limited measurement of $E$-mode polarization up to $\ell \sim 100$, allowing the tightest constraints on the redshift of reionization from the CMB to date.

\section{Acknowledgements}
The CLASS project is supported by the National Science Foundation under Grant Number 0959349. CLASS operates in the Parque Astron\'{o}mico Atacama in northern Chile under the auspices of the Comisi\'{o}n Nacional de Investigaci\'{o}n Cient\'{i}fica y Tecnol\'{o}gica de Chile (CONICYT). A NASA ROSES/APRA grant provided support for the detector technology development. Detector development work at JHU was funded by NASA grant number NNX14AB76A. T.E.-H. is supported by a National Science Foundation Astronomy and Astrophysics Postdoctoral Fellowship.

%\bibliography{class_spie} % Bibliography data in class_spie.bib

\begin{thebibliography}{10}

\bibitem{2013ApJS..208...20B}
{Bennett}, C.~L., {Larson}, D., {Weiland}, J.~L., {Jarosik}, N., {Hinshaw}, G.,
  {Odegard}, N., {Smith}, K.~M., {Hill}, R.~S., {Gold}, B., {Halpern}, M.,
  {Komatsu}, E., {Nolta}, M.~R., {Page}, L., {Spergel}, D.~N., {Wollack}, E.,
  {Dunkley}, J., {Kogut}, A., {Limon}, M., {Meyer}, S.~S., {Tucker}, G.~S., and
  {Wright}, E.~L., ``{Nine-year Wilkinson Microwave Anisotropy Probe (WMAP)
  Observations: Final Maps and Results},'' {\em \apjs}~{\bf 208},  20 (Oct.
  2013).

\bibitem{2013arXiv1303.5076P}
{Planck Collaboration}, {Ade}, P.~A.~R., {Aghanim}, N., {Armitage-Caplan}, C.,
  {Arnaud}, M., {Ashdown}, M., {Atrio-Barandela}, F., {Aumont}, J.,
  {Baccigalupi}, C., {Banday}, A.~J., and et~al., ``{Planck 2013 results. XVI.
  Cosmological parameters},'' {\em ArXiv e-prints}  (Mar. 2013).

\bibitem{2014JCAP...04..014D}
{Das}, S., {Louis}, T., {Nolta}, M.~R., {Addison}, G.~E., {Battistelli}, E.~S.,
  {Bond}, J.~R., {Calabrese}, E., {Crichton}, D., {Devlin}, M.~J., {Dicker},
  S., {Dunkley}, J., {D{\"u}nner}, R., {Fowler}, J.~W., {Gralla}, M., {Hajian},
  A., {Halpern}, M., {Hasselfield}, M., {Hilton}, M., {Hincks}, A.~D.,
  {Hlozek}, R., {Huffenberger}, K.~M., {Hughes}, J.~P., {Irwin}, K.~D.,
  {Kosowsky}, A., {Lupton}, R.~H., {Marriage}, T.~A., {Marsden}, D.,
  {Menanteau}, F., {Moodley}, K., {Niemack}, M.~D., {Page}, L.~A., {Partridge},
  B., {Reese}, E.~D., {Schmitt}, B.~L., {Sehgal}, N., {Sherwin}, B.~D.,
  {Sievers}, J.~L., {Spergel}, D.~N., {Staggs}, S.~T., {Swetz}, D.~S.,
  {Switzer}, E.~R., {Thornton}, R., {Trac}, H., and {Wollack}, E., ``{The
  Atacama Cosmology Telescope: temperature and gravitational lensing power
  spectrum measurements from three seasons of data},'' {\em \jcap}~{\bf 4},  14
  (Apr. 2014).

\bibitem{2012ApJ...755...70R}
{Reichardt}, C.~L., {Shaw}, L., {Zahn}, O., {Aird}, K.~A., {Benson}, B.~A.,
  {Bleem}, L.~E., {Carlstrom}, J.~E., {Chang}, C.~L., {Cho}, H.~M., {Crawford},
  T.~M., {Crites}, A.~T., {de Haan}, T., {Dobbs}, M.~A., {Dudley}, J.,
  {George}, E.~M., {Halverson}, N.~W., {Holder}, G.~P., {Holzapfel}, W.~L.,
  {Hoover}, S., {Hou}, Z., {Hrubes}, J.~D., {Joy}, M., {Keisler}, R., {Knox},
  L., {Lee}, A.~T., {Leitch}, E.~M., {Lueker}, M., {Luong-Van}, D., {McMahon},
  J.~J., {Mehl}, J., {Meyer}, S.~S., {Millea}, M., {Mohr}, J.~J., {Montroy},
  T.~E., {Natoli}, T., {Padin}, S., {Plagge}, T., {Pryke}, C., {Ruhl}, J.~E.,
  {Schaffer}, K.~K., {Shirokoff}, E., {Spieler}, H.~G., {Staniszewski}, Z.,
  {Stark}, A.~A., {Story}, K., {van Engelen}, A., {Vanderlinde}, K., {Vieira},
  J.~D., and {Williamson}, R., ``{A Measurement of Secondary Cosmic Microwave
  Background Anisotropies with Two Years of South Pole Telescope
  Observations},'' {\em \apj}~{\bf 755},  70 (Aug. 2012).

\bibitem{2014PRL1403.3985B}
{BICEP2 Collaboration}, {Ade}, P.~A.~R., {Aikin}, R.~W., {Barkats}, D.,
  {Benton}, S.~J., {Bischoff}, C.~A., {Bock}, J.~J., {Brevik}, J.~A., {Buder},
  I., {Bullock}, E., {Dowell}, C.~D., {Duband}, L., {Filippini}, J.~P.,
  {Fliescher}, S., {Golwala}, S.~R., {Halpern}, M., {Hasselfield}, M.,
  {Hildebrandt}, S.~R., {Hilton}, G.~C., {Hristov}, V.~V., {Irwin}, K.~D.,
  {Karkare}, K.~S., {Kaufman}, J.~P., {Keating}, B.~G., {Kernasovskiy}, S.~A.,
  {Kovac}, J.~M., {Kuo}, C.~L., {Leitch}, E.~M., {Lueker}, M., {Mason}, P.,
  {Netterfield}, C.~B., {Nguyen}, H.~T., {O'Brient}, R., {Ogburn}, IV, R.~W.,
  {Orlando}, A., {Pryke}, C., {Reintsema}, C.~D., {Richter}, S., {Schwarz}, R.,
  {Sheehy}, C.~D., {Staniszewski}, Z.~K., {Sudiwala}, R.~V., {Teply}, G.~P.,
  {Tolan}, J.~E., {Turner}, A.~D., {Vieregg}, A.~G., {Wong}, C.~L., and {Yoon},
  K.~W., ``{Detection Of B-mode Polarization at Degree Angular Scales by
  BICEP2},'' {\em Phys. Rev. Letters}~{\bf 112},  241101 (2014).

\bibitem{2014PhRvD..89d7501K}
{Krauss}, L.~M. and {Wilczek}, F., ``{Using cosmology to establish the
  quantization of gravity},'' {\em \prd}~{\bf 89},  047501 (Feb. 2014).

\bibitem{1999ApJ...512..511M}
{Mather}, J.~C., {Fixsen}, D.~J., {Shafer}, R.~A., {Mosier}, C., and
  {Wilkinson}, D.~T., ``{Calibrator Design for the COBE Far-Infrared Absolute
  Spectrophotometer (FIRAS)},'' {\em \apj}~{\bf 512},  511--520 (Feb. 1999).

\bibitem{2010ApJ...711.1123C}
{Chiang}, H.~C., {Ade}, P.~A.~R., {Barkats}, D., {Battle}, J.~O., {Bierman},
  E.~M., {Bock}, J.~J., {Dowell}, C.~D., {Duband}, L., {Hivon}, E.~F.,
  {Holzapfel}, W.~L., {Hristov}, V.~V., {Jones}, W.~C., {Keating}, B.~G.,
  {Kovac}, J.~M., {Kuo}, C.~L., {Lange}, A.~E., {Leitch}, E.~M., {Mason},
  P.~V., {Matsumura}, T., {Nguyen}, H.~T., {Ponthieu}, N., {Pryke}, C.,
  {Richter}, S., {Rocha}, G., {Sheehy}, C., {Takahashi}, Y.~D., {Tolan}, J.~E.,
  and {Yoon}, K.~W., ``{Measurement of Cosmic Microwave Background Polarization
  Power Spectra from Two Years of BICEP Data},'' {\em \apj}~{\bf 711},
  1123--1140 (Mar. 2010).

\bibitem{2006ApJ...647..813M}
{Montroy}, T.~E., {Ade}, P.~A.~R., {Bock}, J.~J., {Bond}, J.~R., {Borrill}, J.,
  {Boscaleri}, A., {Cabella}, P., {Contaldi}, C.~R., {Crill}, B.~P., {de
  Bernardis}, P., {De Gasperis}, G., {de Oliveira-Costa}, A., {De Troia}, G.,
  {di Stefano}, G., {Hivon}, E., {Jaffe}, A.~H., {Kisner}, T.~S., {Jones},
  W.~C., {Lange}, A.~E., {Masi}, S., {Mauskopf}, P.~D., {MacTavish}, C.~J.,
  {Melchiorri}, A., {Natoli}, P., {Netterfield}, C.~B., {Pascale}, E.,
  {Piacentini}, F., {Pogosyan}, D., {Polenta}, G., {Prunet}, S., {Ricciardi},
  S., {Romeo}, G., {Ruhl}, J.~E., {Santini}, P., {Tegmark}, M., {Veneziani},
  M., and {Vittorio}, N., ``{A Measurement of the CMB $\langle EE \rangle$
  Spectrum from the 2003 Flight of BOOMERANG},'' {\em \apj}~{\bf 647},
  813--822 (Aug. 2006).

\bibitem{2008ApJ...684..771B}
{Bischoff}, C., {Hyatt}, L., {McMahon}, J.~J., {Nixon}, G.~W., {Samtleben}, D.,
  {Smith}, K.~M., {Vanderlinde}, K., {Barkats}, D., {Farese}, P., {Gaier}, T.,
  {Gundersen}, J.~O., {Hedman}, M.~M., {Staggs}, S.~T., and {Winstein}, B.,
  ``{New Measurements of Fine-Scale CMB Polarization Power Spectra from CAPMAP
  at Both 40 and 90 GHz},'' {\em \apj}~{\bf 684},  771--789 (Sept. 2008).

\bibitem{2007ApJ...660..976S}
{Sievers}, J.~L., {Achermann}, C., {Bond}, J.~R., {Bronfman}, L., {Bustos}, R.,
  {Contaldi}, C.~R., {Dickinson}, C., {Ferreira}, P.~G., {Jones}, M.~E.,
  {Lewis}, A.~M., {Mason}, B.~S., {May}, J., {Myers}, S.~T., {Oyarce}, N.,
  {Padin}, S., {Pearson}, T.~J., {Pospieszalski}, M., {Readhead}, A.~C.~S.,
  {Reeves}, R., {Taylor}, A.~C., and {Torres}, S., ``{Implications of the
  Cosmic Background Imager Polarization Data},'' {\em \apj}~{\bf 660},
  976--987 (May 2007).

\bibitem{2005ApJ...624...10L}
{Leitch}, E.~M., {Kovac}, J.~M., {Halverson}, N.~W., {Carlstrom}, J.~E.,
  {Pryke}, C., and {Smith}, M.~W.~E., ``{Degree Angular Scale Interferometer 3
  Year Cosmic Microwave Background Polarization Results},'' {\em \apj}~{\bf
  624},  10--20 (May 2005).

\bibitem{2009ApJ...705..978B}
{Brown}, M.~L., {Ade}, P., {Bock}, J., {Bowden}, M., {Cahill}, G., {Castro},
  P.~G., {Church}, S., {Culverhouse}, T., {Friedman}, R.~B., {Ganga}, K.,
  {Gear}, W.~K., {Gupta}, S., {Hinderks}, J., {Kovac}, J., {Lange}, A.~E.,
  {Leitch}, E., {Melhuish}, S.~J., {Memari}, Y., {Murphy}, J.~A., {Orlando},
  A., {O'Sullivan}, C., {Piccirillo}, L., {Pryke}, C., {Rajguru}, N.,
  {Rusholme}, B., {Schwarz}, R., {Taylor}, A.~N., {Thompson}, K.~L., {Turner},
  A.~H., {Wu}, E.~Y.~S., {Zemcov}, M., and {QUa D Collaboration}, ``{Improved
  Measurements of the Temperature and Polarization of the Cosmic Microwave
  Background from QUaD},'' {\em \apj}~{\bf 705},  978--999 (Nov. 2009).

\bibitem{2007ApJ...665...55W}
{Wu}, J.~H.~P., {Zuntz}, J., {Abroe}, M.~E., {Ade}, P.~A.~R., {Bock}, J.,
  {Borrill}, J., {Collins}, J., {Hanany}, S., {Jaffe}, A.~H., {Johnson}, B.~R.,
  {Jones}, T., {Lee}, A.~T., {Matsumura}, T., {Rabii}, B., {Renbarger}, T.,
  {Richards}, P.~L., {Smoot}, G.~F., {Stompor}, R., {Tran}, H.~T., and
  {Winant}, C.~D., ``{MAXIPOL: Data Analysis and Results},'' {\em \apj}~{\bf
  665},  55--66 (Aug. 2007).

\bibitem{2012arXiv1207.5034Q}
{QUIET Collaboration}, {Araujo}, D., {Bischoff}, C., {Brizius}, A., {Buder},
  I., {Chinone}, Y., {Cleary}, K., {Dumoulin}, R.~N., {Kusaka}, A., {Monsalve},
  R., {N{\ae}ss}, S.~K., {Newburgh}, L.~B., {Reeves}, R., {Wehus}, I.~K.,
  {Zwart}, J.~T.~L., {Bronfman}, L., {Bustos}, R., {Church}, S.~E.,
  {Dickinson}, C., {Eriksen}, H.~K., {Gaier}, T., {Gundersen}, J.~O.,
  {Hasegawa}, M., {Hazumi}, M., {Huffenberger}, K.~M., {Ishidoshiro}, K.,
  {Jones}, M.~E., {Kangaslahti}, P., {Kapner}, D.~J., {Kubik}, D., {Lawrence},
  C.~R., {Limon}, M., {McMahon}, J.~J., {Miller}, A.~D., {Nagai}, M., {Nguyen},
  H., {Nixon}, G., {Pearson}, T.~J., {Piccirillo}, L., {Radford}, S.~J.~E.,
  {Readhead}, A.~C.~S., {Richards}, J.~L., {Samtleben}, D., {Seiffert}, M.,
  {Shepherd}, M.~C., {Smith}, K.~M., {Staggs}, S.~T., {Tajima}, O., {Thompson},
  K.~L., {Vanderlinde}, K., and {Williamson}, R., ``{Second Season QUIET
  Observations: Measurements of the CMB Polarization Power Spectrum at 95
  GHz},'' {\em ArXiv e-prints}  (July 2012).

\bibitem{2013arXiv1312.6646P}
{POLARBEAR Collaboration}, {Ade}, P.~A.~R., {Akiba}, Y., {Anthony}, A.~E.,
  {Arnold}, K., {Atlas}, M., {Barron}, D., {Boettger}, D., {Borrill}, J.,
  {Chapman}, S., {Chinone}, Y., {Dobbs}, M., {Elleflot}, T., {Errard}, J.,
  {Fabbian}, G., {Feng}, C., {Flanigan}, D., {Gilbert}, A., {Grainger}, W.,
  {Halverson}, N.~W., {Hasegawa}, M., {Hattori}, K., {Hazumi}, M., {Holzapfel},
  W.~L., {Hori}, Y., {Howard}, J., {Hyland}, P., {Inoue}, Y., {Jaehnig}, G.~C.,
  {Jaffe}, A., {Keating}, B., {Kermish}, Z., {Keskitalo}, R., {Kisner}, T., {Le
  Jeune}, M., {Lee}, A.~T., {Linder}, E., {Leitch}, E.~M., {Lungu}, M.,
  {Matsuda}, F., {Matsumura}, T., {Meng}, X., {Miller}, N.~J., {Morii}, H.,
  {Moyerman}, S., {Myers}, M.~J., {Navaroli}, M., {Nishino}, H., {Paar}, H.,
  {Peloton}, J., {Quealy}, E., {Rebeiz}, G., {Reichardt}, C.~L., {Richards},
  P.~L., {Ross}, C., {Schanning}, I., {Schenck}, D.~E., {Sherwin}, B.,
  {Shimizu}, A., {Shimmin}, C., {Shimon}, M., {Siritanasak}, P., {Smecher}, G.,
  {Spieler}, H., {Stebor}, N., {Steinbach}, B., {Stompor}, R., {Suzuki}, A.,
  {Takakura}, S., {Tomaru}, T., {Wilson}, B., {Yadav}, A., and {Zahn}, O.,
  ``{Measurement of the Cosmic Microwave Background Polarization Lensing Power
  Spectrum with the POLARBEAR experiment},'' {\em ArXiv e-prints}  (Dec. 2013).

\bibitem{2013PhRvL.111n1301H}
{Hanson}, D., {Hoover}, S., {Crites}, A., {Ade}, P.~A.~R., {Aird}, K.~A.,
  {Austermann}, J.~E., {Beall}, J.~A., {Bender}, A.~N., {Benson}, B.~A.,
  {Bleem}, L.~E., {Bock}, J.~J., {Carlstrom}, J.~E., {Chang}, C.~L., {Chiang},
  H.~C., {Cho}, H.-M., {Conley}, A., {Crawford}, T.~M., {de Haan}, T., {Dobbs},
  M.~A., {Everett}, W., {Gallicchio}, J., {Gao}, J., {George}, E.~M.,
  {Halverson}, N.~W., {Harrington}, N., {Henning}, J.~W., {Hilton}, G.~C.,
  {Holder}, G.~P., {Holzapfel}, W.~L., {Hrubes}, J.~D., {Huang}, N., {Hubmayr},
  J., {Irwin}, K.~D., {Keisler}, R., {Knox}, L., {Lee}, A.~T., {Leitch}, E.,
  {Li}, D., {Liang}, C., {Luong-Van}, D., {Marsden}, G., {McMahon}, J.~J.,
  {Mehl}, J., {Meyer}, S.~S., {Mocanu}, L., {Montroy}, T.~E., {Natoli}, T.,
  {Nibarger}, J.~P., {Novosad}, V., {Padin}, S., {Pryke}, C., {Reichardt},
  C.~L., {Ruhl}, J.~E., {Saliwanchik}, B.~R., {Sayre}, J.~T., {Schaffer},
  K.~K., {Schulz}, B., {Smecher}, G., {Stark}, A.~A., {Story}, K.~T., {Tucker},
  C., {Vanderlinde}, K., {Vieira}, J.~D., {Viero}, M.~P., {Wang}, G.,
  {Yefremenko}, V., {Zahn}, O., and {Zemcov}, M., ``{Detection of B-Mode
  Polarization in the Cosmic Microwave Background with Data from the South Pole
  Telescope},'' {\em Physical Review Letters}~{\bf 111},  141301 (Oct. 2013).

\bibitem{2014arXiv1403.2369T}
{The POLARBEAR Collaboration}, {Ade}, P.~A.~R., {Akiba}, Y., {Anthony}, A.~E.,
  {Arnold}, K., {Atlas}, M., {Barron}, D., {Boettger}, D., {Borrill}, J.,
  {Chapman}, S., {Chinone}, Y., {Dobbs}, M., {Elleflot}, T., {Errard}, J.,
  {Fabbian}, G., {Feng}, C., {Flanigan}, D., {Gilbert}, A., {Grainger}, W.,
  {Halverson}, N.~W., {Hasegawa}, M., {Hattori}, K., {Hazumi}, M., {Holzapfel},
  W.~L., {Hori}, Y., {Howard}, J., {Hyland}, P., {Inoue}, Y., {Jaehnig}, G.~C.,
  {Jaffe}, A.~H., {Keating}, B., {Kermish}, Z., {Keskitalo}, R., {Kisner}, T.,
  {Le Jeune}, M., {Lee}, A.~T., {Leitch}, E.~M., {Linder}, E., {Lungu}, M.,
  {Matsuda}, F., {Matsumura}, T., {Meng}, X., {Miller}, N.~J., {Morii}, H.,
  {Moyerman}, S., {Myers}, M.~J., {Navaroli}, M., {Nishino}, H., {Paar}, H.,
  {Peloton}, J., {Poletti}, D., {Quealy}, E., {Rebeiz}, G., {Reichardt}, C.~L.,
  {Richards}, P.~L., {Ross}, C., {Schanning}, I., {Schenck}, D.~E., {Sherwin},
  B.~D., {Shimizu}, A., {Shimmin}, C., {Shimon}, M., {Siritanasak}, P.,
  {Smecher}, G., {Spieler}, H., {Stebor}, N., {Steinbach}, B., {Stompor}, R.,
  {Suzuki}, A., {Takakura}, S., {Tomaru}, T., {Wilson}, B., {Yadav}, A., and
  {Zahn}, O., ``{A Measurement of the Cosmic Microwave Background B-Mode
  Polarization Power Spectrum at Sub-Degree Scales with POLARBEAR},'' {\em
  ArXiv e-prints}  (Mar. 2014).

\bibitem{2013arXiv1312.1300P}
{Planck Collaboration}, {Abergel}, A., {Ade}, P.~A.~R., {Aghanim}, N., {Alina},
  D., {Alves}, M.~I.~R., {Aniano}, G., {Armitage-Caplan}, C., {Arnaud}, M.,
  {Ashdown}, M., and et~al., ``{Planck 2013 results. XI. All-sky model of
  thermal dust emission},'' {\em ArXiv e-prints}  (Dec. 2013).

\bibitem{2014arXiv1404.5323F}
{Fuskeland}, U., {Wehus}, I.~K., {Eriksen}, H.~K., and {N{\ae}ss}, S.~K.,
  ``{Spatial variations in the spectral index of polarized synchrotron emission
  in the 9-year WMAP sky maps},'' {\em ArXiv e-prints}  (Apr. 2014).

\bibitem{2007ApJ...665..355K}
{Kogut}, A., {Dunkley}, J., {Bennett}, C.~L., {Dor{\'e}}, O., {Gold}, B.,
  {Halpern}, M., {Hinshaw}, G., {Jarosik}, N., {Komatsu}, E., {Nolta}, M.~R.,
  {Odegard}, N., {Page}, L., {Spergel}, D.~N., {Tucker}, G.~S., {Weiland},
  J.~L., {Wollack}, E., and {Wright}, E.~L., ``{Three-Year Wilkinson Microwave
  Anisotropy Probe (WMAP) Observations: Foreground Polarization},'' {\em
  \apj}~{\bf 665},  355--362 (Aug. 2007).

\bibitem{2014arXiv1405.0871P}
{Planck Collaboration}, {Ade}, P.~A.~R., {Aghanim}, N., {Alina}, D., {Alves},
  M.~I.~R., {Armitage-Caplan}, C., {Arnaud}, M., {Arzoumanian}, D., {Ashdown},
  M., {Atrio-Barandela}, F., and et~al., ``{Planck intermediate results. XIX.
  An overview of the polarized thermal emission from Galactic dust},'' {\em
  ArXiv e-prints}  (May 2014).

\bibitem{2000ApJ...543..787L}
{Lay}, O.~P. and {Halverson}, N.~W., ``{The Impact of Atmospheric Fluctuations
  on Degree-Scale Imaging of the Cosmic Microwave Background},'' {\em
  \apj}~{\bf 543},  787--798 (Nov. 2000).

\bibitem{2010ApJ...708.1674S}
{Sayers}, J., {Golwala}, S.~R., {Ade}, P.~A.~R., {Aguirre}, J.~E., {Bock},
  J.~J., {Edgington}, S.~F., {Glenn}, J., {Goldin}, A., {Haig}, D., {Lange},
  A.~E., {Laurent}, G.~T., {Mauskopf}, P.~D., {Nguyen}, H.~T., {Rossinot}, P.,
  and {Schlaerth}, J., ``{Studies of Millimeter-wave Atmospheric Noise above
  Mauna Kea},'' {\em \apj}~{\bf 708},  1674--1691 (Jan. 2010).

\bibitem{2005ApJ...622.1343B}
{Bussmann}, R.~S., {Holzapfel}, W.~L., and {Kuo}, C.~L., ``{Millimeter
  Wavelength Brightness Fluctuations of the Atmosphere above the South Pole},''
  {\em \apj}~{\bf 622},  1343--1355 (Apr. 2005).

\bibitem{Pardo_atmosphere_atacama}
{Pardo}, J.~R., {Cernicharo}, J., and {Serabyn}, R., ``{Atmospheric
  Transmission at Microwaves (ATM): An Improved Model for
  Millimeter/Submillimeter Applications},'' {\em IEEE T Antenn Propag}  (Dec.
  2001).

\bibitem{Rostem2012SPIE}
{Rostem}, K., {Bennett}, C.~L., {Chuss}, D.~T., {Costen}, E., {Crowe}, E.,
  {Denis}, K.~L., {Eimer}, J.~R., {Lourie}, N., {Essinger-Hileman}, T.,
  {Marriage}, T., {Moseley}, S.~H., {Stevenson}, R., {Towner}, D.~W.,
  {Voellmer}, G., {Wollack}, E.~J., and {Zeng}, L., ``{Detector Architecture of
  the Cosmology Large Angular Scale Surveyor},'' in [{\em Society of
  Photo-Optical Instrumentation Engineers (SPIE) Conference
  Series}{\nolinebreak\hspace{0.1em}]},  {\em Society of Photo-Optical
  Instrumentation Engineers (SPIE) Conference Series} {\bf 8452} (2012).

\bibitem{2011ApJ...737...78K}
{Katayama}, N. and {Komatsu}, E., ``{Simple Foreground Cleaning Algorithm for
  Detecting Primordial B-mode Polarization of the Cosmic Microwave
  Background},'' {\em \apj}~{\bf 737},  78 (Aug. 2011).

\bibitem{2009MNRAS.397.1355E}
{Efstathiou}, G., {Gratton}, S., and {Paci}, F., ``{Impact of Galactic
  polarized emission on B-mode detection at low multipoles},'' {\em
  \mnras}~{\bf 397},  1355--1373 (Aug. 2009).

\bibitem{watts_class_foregrounds2014}
{Watts}, D. et~al., ``{Foreground-Cleaning for the Cosmology Large Angular
  Scale Surveyor},''

\bibitem{2009AIPC.1185..371D}
{Denis}, K.~L., {Cao}, N.~T., {Chuss}, D.~T., {Eimer}, J., {Hinderks}, J.~R.,
  {Hsieh}, W.-T., {Moseley}, S.~H., {Stevenson}, T.~R., {Talley}, D.~J.,
  {U.-Yen}, K., and {Wollack}, E.~J., ``{Fabrication of an Antenna-Coupled
  Bolometer for Cosmic Microwave Background Polarimetry},'' in [{\em American
  Institute of Physics Conference Series}{\nolinebreak\hspace{0.1em}]},
  {Young}, B., {Cabrera}, B., and {Miller}, A., eds., {\em American Institute
  of Physics Conference Series} {\bf 1185},  371--374 (Dec. 2009).

\bibitem{appel_spie_2014}
{Appel}, J.~W., {Ali}, A., {Amiri}, M., {Araujo}, D., {Bennett}, C.~L.,
  {Boone}, F., {Chan}, M., {Cho}, H., {Chuss}, D.~T., {Colazo}, F., {Crowe},
  E., {Denis}, K., {D{\"u}nner}, R., {Eimer}, J., {Essinger-Hileman}, T.,
  {Gothe}, D., {Halpern}, M., {Harrington}, K., {Hilton}, G., {Hinshaw}, G.~F.,
  {Huang}, C., {Irwin}, K., {Jones}, G., {Karakla}, J., {Kogut}, A.~J.,
  {Larson}, D., {Limon}, M., {Lowry}, L., {Marriage}, T., {Mehrle}, N.,
  {Miller}, A.~D., {Miller}, N., {Moseley}, S.~H., {Novak}, G., {Reintsema},
  C., {Rostem}, K., {Stevenson}, T., {Towner}, D., {U-Yen}, K., {Wagner},
  E.and~{Watts}, D., {Wollack}, E., {Xu}, Z., and {Zeng}, L., ``{The cosmology
  large angular scale surveyor (CLASS): 38 GHz detector array of bolometric
  polarimeters},'' in [{\em Millimeter, Submillimeter, and Far-Infrared
  Detectors and Instrumentation for Astronomy
  VII}{\nolinebreak\hspace{0.1em}]},  {\em SPIE} {\bf 915354} (June 2014).

\bibitem{rostem_spie_2014}
{Rostem}, K., {Colazo}, F., {Chuss}, D.~T., {Crowe}, E., {Denis}, K.~L.,
  {Moseley}, S.~H., {Stevenson}, T.~R., {Towner}, D.~W., {U-Yen}, K.,
  {Wollack}, E.~J., {Ali}, A., {Appel}, J.~W., {Bennett}, C.~L.,
  {Essinger-Hileman}, T., and {Marriage}, T.~A., ``{Scalable background-limited
  polarization-sensitive detectors for mm-wave applications},'' in [{\em
  Millimeter, Submillimeter, and Far-Infrared Detectors and Instrumentation for
  Astronomy VII}{\nolinebreak\hspace{0.1em}]},  {\em SPIE} {\bf 915311} (2014).

\bibitem{2008JLTP..151..908B}
{Battistelli}, E.~S., {Amiri}, M., {Burger}, B., {Halpern}, M., {Knotek}, S.,
  {Ellis}, M., {Gao}, X., {Kelly}, D., {Macintosh}, M., {Irwin}, K., and
  {Reintsema}, C., ``{Functional Description of Read-out Electronics for
  Time-Domain Multiplexed Bolometers for Millimeter and Sub-millimeter
  Astronomy},'' {\em Journal of Low Temperature Physics}~{\bf 151},  908--914
  (May 2008).

\bibitem{2014arXiv1403.4302B}
{BICEP2 Collaboration}, {Ade}, P.~A.~R., {Aikin}, R.~W., {Amiri}, M.,
  {Barkats}, D., {Benton}, S.~J., {Bischoff}, C.~A., {Bock}, J.~J., {Brevik},
  J.~A., {Buder}, I., {Bullock}, E., {Davis}, G., {Dowell}, C.~D., {Duband},
  L., {Filippini}, J.~P., {Fliescher}, S., {Golwala}, S.~R., {Halpern}, M.,
  {Hasselfield}, M., {Hildebrandt}, S.~R., {Hilton}, G.~C., {Hristov}, V.~V.,
  {Irwin}, K.~D., {Karkare}, K.~S., {Kaufman}, J.~P., {Keating}, B.~G.,
  {Kernasovskiy}, S.~A., {Kovac}, J.~M., {Kuo}, C.~L., {Leitch}, E.~M.,
  {Llombart}, N., {Lueker}, M., {Netterfield}, C.~B., {Nguyen}, H.~T.,
  {O'Brient}, R., {Ogburn}, IV, R.~W., {Orlando}, A., {Pryke}, C., {Reintsema},
  C.~D., {Richter}, S., {Schwarz}, R., {Sheehy}, C.~D., {Staniszewski}, Z.~K.,
  {Story}, K.~T., {Sudiwala}, R.~V., {Teply}, G.~P., {Tolan}, J.~E., {Turner},
  A.~D., {Vieregg}, A.~G., {Wilson}, P., {Wong}, C.~L., and {Yoon}, K.~W.,
  ``{BICEP2 II: Experiment and Three-Year Data Set},'' {\em ArXiv e-prints}
  (Mar. 2014).

\bibitem{2009AIPC.1185..494E}
{Essinger-Hileman}, T., {Appel}, J.~W., {Beal}, J.~A., {Cho}, H.~M., {Fowler},
  J., {Halpern}, M., {Hasselfield}, M., {Irwin}, K.~D., {Marriage}, T.~A.,
  {Niemack}, M.~D., {Page}, L., {Parker}, L.~P., {Pufu}, S., {Staggs}, S.~T.,
  {Stryzak}, O., {Visnjic}, C., {Yoon}, K.~W., and {Zhao}, Y., ``{The Atacama
  B-Mode Search: CMB Polarimetry with Transition-Edge-Sensor Bolometers},'' in
  [{\em American Institute of Physics Conference
  Series}{\nolinebreak\hspace{0.1em}]},  {B.~Young, B.~Cabrera, \& A.~Miller},
  ed., {\em American Institute of Physics Conference Series} {\bf 1185},
  494--497 (Dec. 2009).

\bibitem{2011ApJS..194...41S}
{Swetz}, D.~S., {Ade}, P.~A.~R., {Amiri}, M., {Appel}, J.~W., {Battistelli},
  E.~S., {Burger}, B., {Chervenak}, J., {Devlin}, M.~J., {Dicker}, S.~R.,
  {Doriese}, W.~B., {D{\"u}nner}, R., {Essinger-Hileman}, T., {Fisher}, R.~P.,
  {Fowler}, J.~W., {Halpern}, M., {Hasselfield}, M., {Hilton}, G.~C., {Hincks},
  A.~D., {Irwin}, K.~D., {Jarosik}, N., {Kaul}, M., {Klein}, J., {Lau}, J.~M.,
  {Limon}, M., {Marriage}, T.~A., {Marsden}, D., {Martocci}, K., {Mauskopf},
  P., {Moseley}, H., {Netterfield}, C.~B., {Niemack}, M.~D., {Nolta}, M.~R.,
  {Page}, L.~A., {Parker}, L., {Staggs}, S.~T., {Stryzak}, O., {Switzer},
  E.~R., {Thornton}, R., {Tucker}, C., {Wollack}, E., and {Zhao}, Y.,
  ``{Overview of the Atacama Cosmology Telescope: Receiver, Instrumentation,
  and Telescope Systems},'' {\em \apjs}~{\bf 194},  41 (June 2011).

\bibitem{2013MNRAS.430.2513H}
{Holland}, W.~S., {Bintley}, D., {Chapin}, E.~L., {Chrysostomou}, A., {Davis},
  G.~R., {Dempsey}, J.~T., {Duncan}, W.~D., {Fich}, M., {Friberg}, P.,
  {Halpern}, M., {Irwin}, K.~D., {Jenness}, T., {Kelly}, B.~D., {MacIntosh},
  M.~J., {Robson}, E.~I., {Scott}, D., {Ade}, P.~A.~R., {Atad-Ettedgui}, E.,
  {Berry}, D.~S., {Craig}, S.~C., {Gao}, X., {Gibb}, A.~G., {Hilton}, G.~C.,
  {Hollister}, M.~I., {Kycia}, J.~B., {Lunney}, D.~W., {McGregor}, H.,
  {Montgomery}, D., {Parkes}, W., {Tilanus}, R.~P.~J., {Ullom}, J.~N.,
  {Walther}, C.~A., {Walton}, A.~J., {Woodcraft}, A.~L., {Amiri}, M.,
  {Atkinson}, D., {Burger}, B., {Chuter}, T., {Coulson}, I.~M., {Doriese},
  W.~B., {Dunare}, C., {Economou}, F., {Niemack}, M.~D., {Parsons}, H.~A.~L.,
  {Reintsema}, C.~D., {Sibthorpe}, B., {Smail}, I., {Sudiwala}, R., and
  {Thomas}, H.~S., ``{SCUBA-2: the 10 000 pixel bolometer camera on the James
  Clerk Maxwell Telescope},'' {\em \mnras}~{\bf 430},  2513--2533 (Apr. 2013).

\bibitem{Eimer2012SPIE}
{Eimer}, J.~R., {Bennett}, C.~L., {Chuss}, D.~T., {Marriage}, T., {Wollack},
  E.~J., and {Zeng}, L., ``{The Cosmology Large Angular Scale Surveyor (CLASS):
  40 GHz Optical Design},'' in [{\em Society of Photo-Optical Instrumentation
  Engineers (SPIE) Conference Series}{\nolinebreak\hspace{0.1em}]},  {\em
  Society of Photo-Optical Instrumentation Engineers (SPIE) Conference Series}
  {\bf 8452} (2012).

\bibitem{2013ApOpt..52.8747D}
{Datta}, R., {Munson}, C.~D., {Niemack}, M.~D., {McMahon}, J.~J., {Britton},
  J., {Wollack}, E.~J., {Beall}, J., {Devlin}, M.~J., {Fowler}, J., {Gallardo},
  P., {Hubmayr}, J., {Irwin}, K., {Newburgh}, L., {Nibarger}, J.~P., {Page},
  L., {Quijada}, M.~A., {Schmitt}, B.~L., {Staggs}, S.~T., {Thornton}, R., and
  {Zhang}, L., ``{Large-aperture wide-bandwidth antireflection-coated silicon
  lenses for millimeter wavelengths},'' {\em \ao}~{\bf 52},  8747 (Dec. 2013).

\bibitem{2010ITAP...58.1383Z}
{Zeng}, L., {Bennett}, C.~L., {Chuss}, D.~T., and {Wollack}, E.~J., ``{A Low
  Cross-Polarization Smooth-Walled Horn With Improved Bandwidth},'' {\em IEEE
  Transactions on Antennas and Propagation}~{\bf 58},  1383--1387 (Apr. 2010).

\bibitem{1981InfPh..21..225B}
{Birch}, J.~R., {Dromey}, J.~D., and {Lesurf}, J., ``{The optical constants of
  some common low-loss polymers between 4 and 40 cm $^{-1}$},'' {\em Infrared
  Physics}~{\bf 21},  225--228 (July 1981).

\bibitem{1991ITMTT..39..352S}
{Seeger}, K., ``{Microwave measurement of the dielectric constant of
  high-density polyethylene},'' {\em IEEE Transactions on Microwave Theory
  Techniques}~{\bf 39},  352--354 (Feb. 1991).

\bibitem{2012arXiv1203.5019M}
{Morris}, C.~M., {Vald{\'e}s Aguilar}, R., {Stier}, A.~V., and {Armitage},
  N.~P., ``{Polarization modulation time-domain terahertz polarimetry},'' {\em
  ArXiv e-prints}  (Mar. 2012).

\bibitem{2003ApJS..145..413J}
{Jarosik}, N., {Bennett}, C.~L., {Halpern}, M., {Hinshaw}, G., {Kogut}, A.,
  {Limon}, M., {Meyer}, S.~S., {Page}, L., {Pospieszalski}, M., {Spergel},
  D.~N., {Tucker}, G.~S., {Wilkinson}, D.~T., {Wollack}, E., {Wright}, E.~L.,
  and {Zhang}, Z., ``{Design, Implementation, and Testing of the Microwave
  Anisotropy Probe Radiometers},'' {\em \apjs}~{\bf 145},  413--436 (Apr.
  2003).

\bibitem{2004ApJ...610..625F}
{Farese}, P.~C. et~al., ``{COMPASS: An Upper Limit on Cosmic Microwave
  Background Polarization at an Angular Scale of 20'},'' {\em \apj}~{\bf 610},
  625--634 (Aug. 2004).

\bibitem{2005ApJS..159....1B}
{Barkats}, D. et~al., ``{Cosmic Microwave Background Polarimetry Using
  Correlation Receivers with the PIQUE and CAPMAP Experiments},'' {\em
  \apjs}~{\bf 159},  1--26 (July 2005).

\bibitem{2012ApJ...760..145Q}
{QUIET Collaboration} et~al., ``{Second Season QUIET Observations: Measurements
  of the Cosmic Microwave Background Polarization Power Spectrum at 95 GHz},''
  {\em \apj}~{\bf 760},  145 (Dec. 2012).

\bibitem{2010A&A...520A...4B}
{Bersanelli}, M. et~al., ``{Planck pre-launch status: Design and description of
  the Low Frequency Instrument},'' {\em \aap}~{\bf 520},  A4 (Sept. 2010).

\bibitem{2006PhDT........33S}
{Stefanescu}, E., {\em {The Ku-Band Polarization Identifier, a new instrument
  to probe polarized astrophysical radiation at 12--18 GHz}}, PhD thesis,
  University of Miami (2006).

\bibitem{2003PhRvD..68d2002O}
{O'dell}, C.~W., {Keating}, B.~G., {de Oliveira-Costa}, A., {Tegmark}, M., and
  {Timbie}, P.~T., ``{CMB polarization at large angular scales: Data analysis
  of the POLAR experiment},'' {\em \prd}~{\bf 68},  042002 (Aug. 2003).

\bibitem{CBI_instrument_2002}
{Padin}, S. et~al., ``{The Cosmic Background Imager},'' {\em Publ. Astron. Soc.
  Pac.}~{\bf 114},  83--97 (2002).

\bibitem{2009ApJ...694.1664C}
{Chen}, M.-T. et~al., ``{AMiBA: Broadband Heterodyne Cosmic Microwave
  Background Interferometry},'' {\em \apj}~{\bf 694},  1664--1669 (Apr. 2009).

\bibitem{2013ApJ...765...64M}
{Moyerman}, S. et~al., ``{Scientific Verification of Faraday Rotation
  Modulators: Detection of Diffuse Polarized Galactic Emission},'' {\em
  \apj}~{\bf 765},  64 (Mar. 2013).

\bibitem{2014RScI...85b4501K}
{Kusaka}, A., {Essinger-Hileman}, T., {Appel}, J.~W., {Gallardo}, P., {Irwin},
  K.~D., {Jarosik}, N., {Nolta}, M.~R., {Page}, L.~A., {Parker}, L.~P.,
  {Raghunathan}, S., {Sievers}, J.~L., {Simon}, S.~M., {Staggs}, S.~T., and
  {Visnjic}, K., ``{Modulation of cosmic microwave background polarization with
  a warm rapidly rotating half-wave plate on the Atacama B-Mode Search
  instrument},'' {\em Review of Scientific Instruments}~{\bf 85},  024501 (Feb.
  2014).

\bibitem{2012ApOpt..51..197C}
{Chuss}, D.~T., {Wollack}, E.~J., {Henry}, R., {Hui}, H., {Juarez}, A.~J.,
  {Krejny}, M., {Moseley}, S.~H., and {Novak}, G., ``{Properties of a
  variable-delay polarization modulator},'' {\em \ao}~{\bf 51},  197 (Jan.
  2012).

\bibitem{2006ApOpt..45.5107C}
{Chuss}, D.~T., {Wollack}, E.~J., {Moseley}, S.~H., and {Novak}, G.,
  ``{Interferometric polarization control},'' {\em \ao}~{\bf 45},  5107--5117
  (July 2006).

\bibitem{2014arXiv1403.1652C}
{Chuss}, D.~T., {Eimer}, J.~R., {Fixsen}, D.~J., {Hinderks}, J., {Kogut},
  A.~J., {Lazear}, J., {Mirel}, P., {Switzer}, E., {Voellmer}, G.~M., and
  {Wollack}, E.~J., ``{Variable-delay Polarization Modulators for Cryogenic
  Millimeter-wave Applications},'' {\em ArXiv e-prints}  (Mar. 2014).

\bibitem{wollack_photonic_choke_2010}
{Wollack}, E.~J., {U-Yen}, K., and {Chuss}, D.~T., ``{Photonic Choke-Joints for
  Dual-Polarization Waveguides},'' {\em Microwave Symposium Digest, 2010 IEEE
  MMT-S International} ,  177--180 (May 2010).

\end{thebibliography}
%\bibliographystyle{spiebib}

\end{document}